\documentclass[11pt]{article}

\date{}

\PassOptionsToPackage{numbers, compress}{natbib}
\usepackage[utf8]{inputenc} % allow utf-8 input
\usepackage[T1]{fontenc}    % use 8-bit T1 fonts
\usepackage{hyperref}       % hyperlinks
\usepackage{url}            % simple URL typesetting
\usepackage{booktabs}       % professional-quality tables
\usepackage{amsfonts}       % blackboard math symbols
\usepackage{nicefrac}       % compact symbols for 1/2, etc.
\usepackage{microtype}      % microtypography
\usepackage{lipsum}
\usepackage{fancyhdr}       % header
\usepackage{graphicx}       % graphics
\graphicspath{{media/}}     % organize your images and other figures under media/ folder
\usepackage{subcaption}
\usepackage{adjustbox}
\usepackage{amsmath}
\usepackage{multirow}
\usepackage{algorithm}
\usepackage{algpseudocode}
\usepackage{comment}

\usepackage{pifont}
\usepackage[usenames,dvipsnames]{xcolor}

\newcommand*\colourcheck[1]{%
  \expandafter\newcommand\csname #1check\endcsname{\textcolor{#1}{\ding{52}}}%
}
\newcommand*\colourxmark[1]{%
  \expandafter\newcommand\csname #1xmark\endcsname{\textcolor{#1}{\ding{55}}}%
}
\colourcheck{blue}
\colourcheck{OliveGreen}

\colourxmark{BrickRed}

\title{Aero-Nef: Neural Fields for Rapid Aircraft Aerodynamics Simulations
}

\author{
  Giovanni Catalani\textsuperscript{1,2}, Siddhant Agarwal \textsuperscript{1}, Xavier Bertrand \textsuperscript{1}, \\
  Frederic Tost \textsuperscript{1}, Michael Bauerheim \textsuperscript{2}, Joseph Morlier \textsuperscript{2} \\
  \\
  1. Airbus \\
  2. ISAE-Supaero, Toulouse \\
  Corresponding email: \texttt{giovanni.catalani@airbus.com} \\
}

\begin{document}
\maketitle

\begin{abstract}
  This paper presents a methodology to learn surrogate models of steady state fluid dynamics simulations on meshed domains, based on Implicit Neural Representations (INRs). The proposed models can be applied directly to unstructured domains for different flow conditions, handle non-parametric 3D geometric variations, and generalize to unseen shapes at test time.
  The coordinate-based formulation naturally leads to robustness with respect to discretization, allowing an excellent trade-off between computational cost (memory footprint and training time) and accuracy. The method is demonstrated on two industrially relevant applications: a RANS dataset of the two-dimensional compressible flow over a transonic airfoil and a dataset of the surface pressure distribution over 3D wings, including shape, inflow condition, and control surface deflection variations. On the considered test cases, our approach achieves a more than three times lower test error and significantly improves generalization error on unseen geometries compared to state-of-the-art Graph Neural Network architectures. Remarkably, the method can perform inference five order of magnitude faster than the high fidelity solver on the RANS transonic airfoil dataset.
\end{abstract}

\section{Introduction}

Computational Fluid Dynamics (CFD) has become an indispensable tool in modern aircraft design, offering the possibility to perform high-fidelity simulations of the complex physics of an aircraft. Numerical solutions can be leveraged to evaluate aerodynamic performance, structural loads, and handling qualities at a much lower cost than experimental and full-scale flight test campaigns. These simulations primarily solve the Reynolds-averaged Navier-Stokes (RANS) \cite{pope2001turbulent} equations to produce accurate flow field predictions, over the full flight envelope \cite{da2011generation}. However, the high computational cost associated with each simulation restricts their use in the early design stages and time-sensitive environments, such as the aerodynamic load estimation required over thousands of flight conditions and design choices.

Given the prohibitive computational demands of traditional CFD simulations, there is a critical need for efficient surrogate models. These models are designed to approximate complex simulations quickly and accurately, facilitating rapid iterations during the design process. Traditionally, methods like Proper Orthogonal Decomposition (POD) \cite{berkooz1993proper} combined with interpolation techniques (e.g., radial basis functions or Gaussian Processes \cite{chiplunkar2018gaussian}) have been prevalent \cite{hijazi2020data, san2019artificial}. Despite their efficiency and physical interpretability, these models often fall short in scenarios involving strongly non-linear phenomena, particularly at transonic flow conditions, due to their inherent linear nature \cite{lucia2002domain}. Additionally, the fixed geometry and resolution constraints of modal decomposition methods limit their applicability to scenarios involving geometric variations and multi-fidelity data. Attempts to circumvent these drawbacks have focused on operating snapshot clustering \cite{dupuis2018aerodynamic,catalani2022machine}, modifying the POD minimization metric \cite{bui2007goal}, and performing mesh morphing to a common reference mesh \cite{casenave2024mmgp}. Despite the potential of these proposed methods, their applicability can be limited to specific cases, due to the underlying assumptions.  

The recent advances in data-driven modeling, especially deep learning (DL), offer new avenues for constructing surrogate models. Deep learning has demonstrated exceptional capability in extracting and representing complex hierarchical data features across various domains \cite{lecun2015deep,bengio2013representation,vaswani2017attention}. In the context of fluid dynamics, DL methods have been used to perform a variety of tasks: constructing reduced order models \cite{fresca2021comprehensive,eivazi2022towards}, accelerating numerical solvers \cite{ajuria2020towards,ranade2022composable}, identifying turbulence closure models \cite{schmelzer2020discovery} or flow control strategies \cite{rabault2019accelerating,corban2023discovering}, are only some of the most relevant examples.  

Specifically for airfoil and aircraft aerodynamic predictions, Deep Learning applications include modeling aerodynamic coefficients \cite{zahn2021application}, surface pressure distributions \cite{hines2023graph} pressure calibration \cite{bertrand2019wing} and shape optimization \cite{baque2018geodesic,wei2024deepgeo,wei2024diffairfoil}. 
Building a surrogate model of the pressure fields, has a central importance in all of the above tasks, replacing the need for a computationally expensive numerical simulator.
Convolutional Neural Networks (CNNs) have been widely used to build surrogates, thanks to the strong capability of these networks to capture local interactions and the possibility to straightforwardly define multiscale operations. UNET architectures have been applied to predict scalar and vector fields over airfoils and 3D configurations \cite{thuerey2020deep,guo2016convolutional,catalani2023comparative}. Despite these successful attempts, CNNs expect pixel-like input and output data at fixed resolution: CFD solutions are typically defined on unstructured meshes, and interpolation routines have been employed to map the two different types of data representations, with an inevitable loss in performance \cite{catalani2023comparative}. Additionally, the different level of refinement characteristic of numerical simulations, hinders the applicability of CNN to 3D configurations, due to computational constraints.

Geometric deep learning, specifically Graph Neural Networks (GNNs), provide a powerful paradigm to extend CNNs to unstructured domains, exploiting the inductive biases inherent to graph-like data \cite{bronstein2017geometric}.
This flexibility allows GNNs to model complex geometries typical of fluid dynamics simulations more naturally than traditional neural network architectures. Mesh Graph Networks \cite{pfaff2020learning} introduce convolution operations on meshes, involving relative node distances as edge features, and demonstrate good performance for fluid mechanics time-dependent problems defined on a common mesh, with a relatively limited number of nodes. Extension of message passing GNNs to larger graphs, inevitably requires pooling operations, in order to model multiscale behavior of fluid flows \cite{fortunato2022multiscale, gao2019graph,lino2022multi} and long range interactions over the mesh. In a monoscale architecture, a larger number of message passing layers would be needed to cover an equivalent graph region, leading to oversmoothing \cite{rusch2023survey} and excessive computational overhead. The definition of pooling operations is, however, dependent on the mesh topology and the kind of application, as no general downsampling rule fits all cases. Despite their general formulation and success across a wide range of applications, the effectiveness of GNN for surrogate modeling of aerodynamic simulations is still limited. Firstly, GNNs architectures encounter challenges in generalizing across different mesh topologies and levels of discretization: this can be explained by the fact the node connectivity and neighborhood definitions are based on the graph's structural metrics rather than the physical metrics of the domain \cite{li2024geometry}. Refining a mesh over a certain threshold often degrades the performance of the GNN \cite{fortunato2022multiscale}, as the neighborhood of each node shrinks, converging to a single point in the limit of infinite refinement, failing the discretization convergence criterion.
Typical industrial aerodynamic applications in 3D involve largely refined meshes (even surfacic-only meshes can exceed 1 Million nodes per geometry) in order to account for the complex interactions between lifting surfaces and the fluid flow at higher Reynolds Numbers. Training a surrogate model at full resolution is often unfeasible with limited hardware resources: scalability requires the possibility of accurately generating predictions at higher levels of discretization while training at significantly coarser levels. It must be noted that a substantial effort to scale GNN to handle large meshes, in the content of aircraft aerodynamics, has been done in the work of Hines and Berkemeyer \cite{hines2023graph}, as dynamic subsampling has been applied to predict the surface pressure distribution on the single NASA CRM aircraft geometry \cite{vassberg2008development} over a range of flight parameters variations.

Recent advances in the area of Operator Learning have allowed the introduction of discretization invariant architecture, bypassing the limitations of GNN which have been mentioned earlier. Neural Operator, learns the mapping between the infinite-dimensional input and output function spaces, and can by construction produce the value of the output function at any point in the spatial domain. Fourier Neural Operator (FNO) \cite{li2020fourier} has been introduced to solve parametric PDEs, by learning the integral kernel directly in the Fourier Space, leveraging the computational efficiency of Fast Fourier Transforms (FFT) on uniform grids. The Geometry Informed Neural Operator (GINO) \cite{li2024geometry} employs a Graph Neural Network to map unstructured 3D domains to a structured reference latent grid, where Fourier Layers can be efficiently applied.  
Similarly, DeepOnet \cite{lu2019deeponet} learns operators from a sparse set of observations using two subnetworks, but it is constrained to input functions evaluated always in the same location, being poorly suited for problems involving shape variations. Recently, Geom-DeepOnet was introduced \cite{he2024geom}, extending the capabilities of the parent DeepOnet methodology to handle parametric geometry variations in 3D. 

Implicit Neural Representations (INRs), also known as Neural Fields, have emerged as a powerful alternative to classical methods for learning spatial representations of objects such as images, shapes, and 3D scenes through radiance fields \cite{park2019deepsdf,chen2019learning,mildenhall2021nerf}. INRs achieve an impressive level of detail with limited memory requirements by approximating functions at any point in the domain. They take spatial coordinates as input and produce the corresponding function value at those coordinates. Recent breakthroughs in the field, particularly in input encoding techniques like SIREN \cite{sitzmann2020implicit} and Fourier Feature Encoding (FFN) \cite{tancik2020fourier}, have significantly enhanced the ability of INRs to approximate high-frequency functions with unmatched efficiency. SIREN employs periodic activation functions, enabling the representation of complex signals, while FFN leverages a set of predefined frequency components typically sampled from a Gaussian distribution.

However, tuning the hyperparameters of these architectures, especially the frequency components in FFN, is non-trivial. The sampled frequencies can significantly impact the model's underfitting or overfitting behavior \cite{tancik2020fourier}. This aspect is particularly challenging in fluid dynamics datasets, which often exhibit a wide spectrum of frequencies. 
While initially limited to representing single samples, advancements such as latent modulation and meta-learning have extended INRs' capabilities to approximate entire classes of objects and datasets \cite{tancik2021learned, chen2019learning, sitzmann2020implicit, dupont2022data}. Serrano et al. introduced CORAL \cite{serrano2024operator}, a flexible framework for learning initial value problems, modeling PDE dynamics, and building surrogate models on 2D meshes using INRs. This framework demonstrated computational efficiency and the ability to handle geometric variations accurately. The INFINITY model \cite{serrano2023infinity} further specialized this approach for the RANS equations over 2D airfoils in the incompressible regime, achieving state-of-the-art accuracy.

\paragraph{In this study} we present a comprehensive methodology to build surrogate models of steady aerodynamic simulations on meshed domains using Implicit Neural Representations. Our contributions are twofold. Firstly, we introduce the first application of INR-based surrogate models to 3D wing surface meshes, incorporating shape variations and aileron deflections over a wide range of flight conditions. This showcases the potential of our model for real industrial applications. Our methodology is specialized into two main frameworks: an encode-process-decode design suitable for general problems with non-parametric shape variations, and an end-to-end design optimized for scenarios with fixed geometry or parametric shape variations.

Secondly, we propose a Multiscale-INR backbone architecture to optimize spectral convergence, addressing the challenges associated with tuning the frequency hyperparameters of standard Fourier Feature Encoding architectures. This architecture ensures that our model captures the necessary frequency components to accurately represent complex aerodynamic phenomena without extensive hyperparameter tuning.

Our extensive experimental study includes two key datasets: a transonic airfoil dataset with fixed mesh configurations and a 3D wing dataset with shape variations and aileron deflections. We chose the 2D transonic airfoil dataset due to its industrial relevance, as accurately predicting shocks is crucial for aircraft aerodynamics, with cruise speeds typically in the transonic range. Traditional methods like POD often struggle to capture these shock phenomena accurately. The 3D wing dataset presents extensive shape variations, making it particularly challenging for most deep learning architectures. We also test the model's generalization capabilities on unseen shapes. These experiments demonstrate the accuracy, efficiency, and scalability of our proposed methodology. We compare our approach with state-of-the-art Graph Neural Network (GNN) baselines, highlighting several aspects crucial to the method's scalability, such as discretization dependency.

Additionally, we conduct a discretization dependence study to quantify our method's ability to generalize across different levels of mesh refinement. This study underscores the scalability potential of our framework, enabling accurate predictions at higher levels of discretization while training at significantly coarser levels. This capability is particularly important for industrial applications, where high-resolution simulations are computationally prohibitive.

\section{Methodology}
\label{sec:methodology}
\paragraph{Problem Statement}

In this work, we address the problem of finding data-driven solutions to steady-state fluid dynamics simulations discretized on meshes. Let $\Omega \subset \mathbb{R}^d$ denote the physical domain of the geometry under consideration, where $d$ indicates the dimensionality of the spatial domain.

Let  $\mathcal{A} = \mathcal{A}(\Omega; \mathbb{R}^{d_a})$ and $\mathcal{U} = \mathcal{U}(\Omega; \mathbb{R}^{d_u})$ be separable Banach spaces of functions taking values in $\mathbb{R}^{d_a}$ and $\mathbb{R}^{d_u}$ respectively. Each element of $\mathcal{A}$ is a function describing the input geometry (for instance through the Signed Distance Function), the boundary conditions and the inflow conditions. Let $\mathcal{G}: \mathcal{A} \rightarrow \mathcal{U}$ be a non-linear map, serving as the solution operator of the associated partial differential equation (PDE) that describes the physics of the problem.

Numerical methods seek for an approximate solution $u \in \mathcal{U}$ of the governing PDE on a discretized version of the input domain $\Omega$, namely a mesh $\mathcal{M}=(X, \mathcal{E})$, defined by the set of nodes coordinates $X\in \mathbb{R}^{N d}$ and their connectivities $\mathcal{E} \in \{0,1 \}^{N \times N}$ for specific instances of the input parameter vector $\mu \in \mathbb{R}^{d_p}$.

Surrogate models are built to approximate the numerical solution, for different values of the input parameters and different geometrical configurations, typically at a much lower cost than the associated numerical methods. Data driven methods learn this mapping from a dataset of observed input-output pairs:
\begin{equation}
    \mathcal{D} = \{(\mathcal{M}_i, \mu_i), U_i\}_{i=1}^{M}
\end{equation}
where each tuple consists of a mesh (${M}_i$), input parameters ($\mu_i$) typically describing the boundary conditions and inflow conditions, and the corresponding field solution ($U_i \in \mathbb{R}^{N_i d_u}$), evaluated on the mesh, usually with a CFD solver.
Additional scalars, namely lift and drag coefficients, can be considered as the output of the surrogate model, but in the following, we will focus on building parametric models for full fields predictions.

Specifically, we aim to define a methodology to build surrogate models of aerodynamic simulation on meshes, by directly approximating the solution operator of the underlying PDE with a data-driven architecture based on Implicit Neural Representations. It must be highlighted that although the primary objective of a surrogate model is to obtain discretized solutions on a given mesh, reformulating the problem in terms of approximating the continuous solutions operator allows to obtain a flexible and scalable algorithm with respect to discretization, as it will be shown in the following paragraphs.

%INR
\paragraph{Implicit Neural Representations}
Implicit Neural Representations (INRs) are coordinate based neural networks parameterized by weights $\theta \in \mathbb{R}^p$ mapping points $\mathbf{x}$ in Euclidean space (or equivalently on a manifold) to vector or scalar quantities:
\begin{equation}
    f_\theta(\mathbf{x}) : \mathbb{R}^d \rightarrow \mathbb{R}^{d_u}.
\end{equation}
INRs can be thought of as a continuous approximation of an underlying signal, for which a discretely sampled version, is available for learning. Leveraging the continuous nature of INRs, numerous data modalities can be modeled, such as images, shapes, and physical fields without constraints on the type of discretization.
In this work, we employ INRs to learn concise representations of physical solutions and shapes descriptors, that are invariant across levels and types of discretization.
Field predictions can be obtained by querying an INR across multiple coordinate points:

\begin{equation}
    f_\theta(X) = [f_\theta(\mathbf{x_1}),..., f_\theta(\mathbf{x_N})]. 
\end{equation}

The Implicit Neural Architecture is implemented through a Multi-Layer-Perceptron (MLP) which in its most basic form takes as input the spatial coordinates and outputs the value of the physical field at the specific coordinate. This formulation is naturally suited to fit single signals defined over low-dimensional domains.
For machine learning tasks, that require learning classes of objects (eg. a dataset of various PDE solutions, instead of a single element), the INR architecture can be conditioned through a latent representation that encodes sample-specific features.

Hence, in the following we proceed to describe: (i) how to train INRs to handle variations in geometry and input parameters, and (ii) how the proposed architectural components make Neural Fields a powerful paradigm for learning mesh-based simulations.

\begin{figure}[h]
  \centering
  \begin{subfigure}[b]{0.45\linewidth}
    \centering
    \raisebox{0.1\height}{\includegraphics[width=\linewidth]{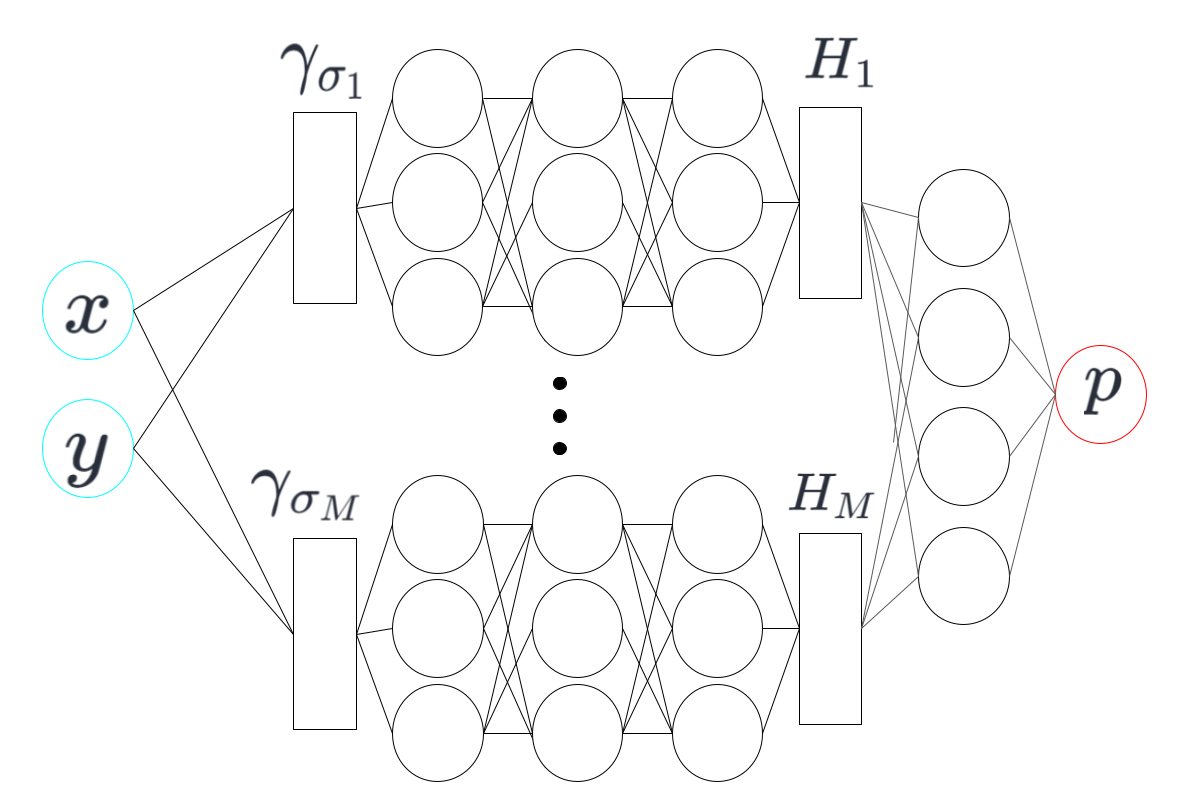}} % Adjust the value to raise the image
  \end{subfigure}
  \hspace{-0.025\linewidth} % Adjust this value to control the spacing between the images
  \begin{subfigure}[b]{0.55\linewidth}
    \centering
    \includegraphics[width=\linewidth]{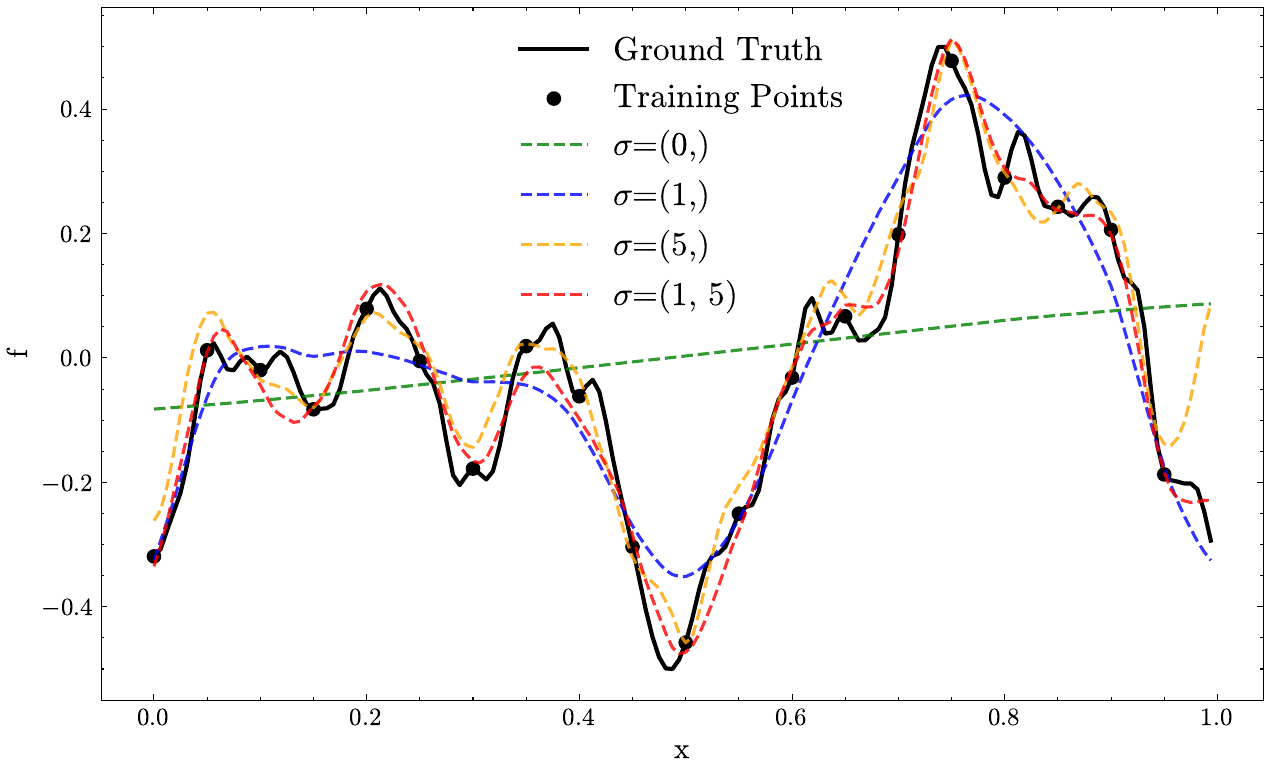} 
  \end{subfigure}
  \caption{\textbf{Left:} Illustration of the multiscale architecture.  \textbf{Right:} Experimental results for fitting a single signal using a 3-layer INR with different Fourier encoding. This experiment can be easily reproduced using the notebook provided in the accompanying Git repository (\url{https://gitlab.isae-supaero.fr/gi.catalani/aero-nepf}).}
  \label{fig:both_subfigures}
\end{figure}

%Fourier Feature Mapping
\paragraph{Input Encoding}
Standard Neural Networks fail to represent high frequency oscillations on low dimensional domains \cite{tancik2020fourier}, being biased towards lower frequency signals. This phenomenon, denominated Spectral Bias, hinders the convergence of vanilla MLP architectures to higher harmonics, leading to blurry reconstructions.
The solutions proposed in the literature employ positional encoding \cite{mildenhall2021nerf} or Fourier Feature encoding \cite{tancik2020fourier} to the inputs. \cite{tancik2020fourier} shows that incorporating these encodings, leads to a stationary Neural Tangent Kernel \cite{jacot2018neural} stationary, thus improving the network convergence to the high frequency component of the signal being approximated. The Fourier Encoding is implemented as follows:
\begin{equation}
    \gamma({\mathbf{x}}) = [a_1 cos(2\pi b_1^T \mathbf{x}),a_1 sin(2\pi b_1^T \mathbf{x}),...,a_n cos(2\pi b_n^T \mathbf{x}), a_n sin(2\pi b_n^T \mathbf{x}) ]
\end{equation}
where the frequency components are sampled
from a zero-centered Gaussian Distribution  $b_i \sim \mathcal{N}(0,\sigma)$.
%Multiscale Encoding
The standard deviation ($\sigma$) of the sampling distribution is a hyper-parameter that can have a significant effect on the underfitting-overfitting behavior of the network \cite{tancik2020fourier,wang2021eigenvector}. In fact, its magnitude determines the range of frequency that the network can represent (this is typically referred to as kernel bandwidth) in the reconstructed signals. Low $\sigma$ values lead to the network acting a low pass filter on the reconstructed signal, causing underfitting behaviour as the network fails to capture the details of the target signal. Larger $\sigma$ values, on the other hand, can lead the network to incorporate high frequency patterns, and produce noisy reconstruction characteristics of overfitting behaviour. 
Attempts to learn the optimal value or infer it from the frequency spectrum of the target signal are limited to simple cases in 1D or single signal scenarios. It can be much more complex to tune this hyperparameter to fit a dataset of CFD simulation, exhibiting multiscale phenomena. To circumvent this issue we propose a multiscale architecture where multiple input encoding are performed using distinct values of $\sigma$, passed through the network leading to intermediate outputs that are concatenated through the final layer.
This approach, inspired by the work of \cite{wang2021eigenvector} in the context of PINNs, appears to simplify other proposed methods in the field of computer vision \cite{hertz2021sape,landgraf2022pins}, based on a progressive increment of $\sigma$ during training. We refer the reader to the corresponding literature for a deeper dive into this topic.
%%image multiscale and equations
The encoding function for each scale \( i \) is defined as:
\begin{equation}
    \gamma_{\sigma_{i}}(\mathbf{x}) = \left[ \sin(2\pi \mathbf{B}^{(i)} \mathbf{x}), \cos(2\pi \mathbf{B}^{(i)} \mathbf{x}) \right]^T, \quad \textit{for } i = 1, 2, \dots, M
\end{equation}
where each element of $\mathbf{B} \in \mathbb{R}^{n \times d}$ is sampled from a Gaussian distribution $\mathcal{N}(0,\sigma_i)$.
The output of the network,  $f_\theta(\mathbf{x})$, is then obtained by passing the concatenated intermediate output at each scale through a final linear layer :
\begin{equation}
     f_\theta(\mathbf{x}) = W_{L+1}  [H_L^{(1)}, H_L^{(2)}, \dots, H_L^{(M)}] + b_{L+1}.
\end{equation}
Note that the hidden layers of the network are shared for all the encoded inputs, and therefore this approach does not considerably increase the parameter count. In Figure \ref{fig:both_subfigures} a schematic illustration of the multiscale architecture is presented. A simple pedagogical experiment is performed by fitting a single 1D signal with large frequency bandwidth: choosing \textit{a priori} one single value of $\sigma$ is not trivial. Larger values of this hyperparameter (eg. $\sigma=5$) can lead to overfitting behavior as high frequency oscillations are present in the reconstructed signal, while smaller sigma values (eg. $\sigma=0$) can result in underfitting. The Multiscale architecture ($\sigma=[1,5]$)leads to better reconstructions, by balancing the two behaviors: this advantage becomes even more relevant when fitting multiple signals in higher dimensions.
%Latent Code Conditioning, Meta-Learning
\paragraph{Conditioning via Modulation}

Conditioning the network to a specific sample is performed through shift modulation of the intermediate layer outputs. The hidden layer outputs are adjusted by a modulation vector $\phi$, which is added to the output of each layer before applying the ReLU activation function. The first layer receives the encoded input \(\gamma_{\sigma_i}(\mathbf{x})\). The composition of layers in the network is described as follows:
\begin{align*}
    H(\mathbf{x}) &= W_{L}(\eta_{L-1} \circ \eta_{L-2}\circ .... \circ  \eta_{1}\circ \gamma_{\sigma}(\mathbf{x})) +b_L\\
    \eta_{l}(\cdot) &=\textit{ReLU}(W_{l}(\cdot) + b_l + \phi_l)
\end{align*}

where $ W_l, b_l \in \theta $ are the weights and biases of the \( l \)-th layer, and \( \phi_l \) is the shift modulation vector for the \( l \)-th layer. Here we omit the index of the scale encoding. The modulation vectors carry the specific information of each sample and allow to use a unique set of global network parameters to account for all parametric or geometry variations. The modulation is computed for each layer. It is a function of a latent embedding ($z$) for the encode-process-decode model or directly of the input parameters ($\mu$) for the end-to-end framework, through a hypernetwork $h_{\psi}$ parameterized by its parameters $\psi$. A detailed description of both frameworks is provided in the following.

\paragraph{End-to-end framework}
Several ways have been proposed to condition Neural Fields to handle classes of shapes or images. In the original DeepSDF model proposed by Park  et al. \cite{park2019deepsdf}, a distinct latent vector is jointly learned at training time together with network parameters to minimize the loss between the network outputs and the Signed Distance Function on a batch of sampled points around a specific shape. Similarly, a hypernetwork $h$ of parameters $\mathbf{\psi}$ can be used to map the input parameters $\mathbf{\mu}$ to the latent modulation vectors $\phi$. In the case where the desired output is a physical field on a fixed geometry, the learning task can be handled end-to-end by minimizing the pointwise distance between the model predictions and the target values across the sampled coordinates and training samples :
\begin{equation}
    \min_{\mathbf{\theta}, \mathbf{\psi}} \mathcal{L} = \sum_{i}^N\sum_{\mathbf{x} \in \mathcal{M}} \ell \left( f_{\mathbf{\theta}}(\mathbf{x}; h_{\mathbf{\psi}}(\mathbf{\mu}_i)), u_i(\mathbf{x}) \right).
\end{equation}
In practice, it is not necessary to provide the full set of sampled coordinate points at each epoch. The model is able to fit the data with a much reduced training resolution, while being able to perform super-resolution at test time. This feature can greatly reduce training time and memory requirements, as only a subset of points can be used for training. In particular, the best strategy to ensure optimal data diversity is to dynamically downsample the training samples at the required training resolution at the beginning of each epoch.

\begin{comment}
We note that another viable model architecture would incorporate two separate modules: an encoding INR, where network parameters and latent codes for training samples are jointly optimized,  and a processor network, mapping the input parameters to the output latent vectors (analogous to the encode-process-decode model described in the following). This solution, despite producing more accurate reconstructions given the optimized latent codes, performs sub-optimally (for the simplified task without geometry encoding) compared to the end-to-end framework as the module mapping the parameters to the output latent codes, induces an additional interpolation error on the final solution. 
Directly learning the latent code, jointly with the global network parameters, enhances the expressiveness of the model and improves its reconstructions: this is beneficial for tasks such as shape reconstruction \cite{park2019deepsdf}. For surrogate modeling, at test time, the latent codes are inferred from the input parameters, and the additional expressiveness can translate into the interpolation error which propagates in the decoding stage. Constraining the latent codes to be the output of a hyper-network, in the end-to-end framework reduces the reconstruction accuracy but does not require separate interpolation stages, providing the lowest overall prediction error.
\end{comment}

\begin{figure}[h!]
    \centering
    \begin{subfigure}[b]{\textwidth}
        \centering
        \includegraphics[width=1\linewidth]{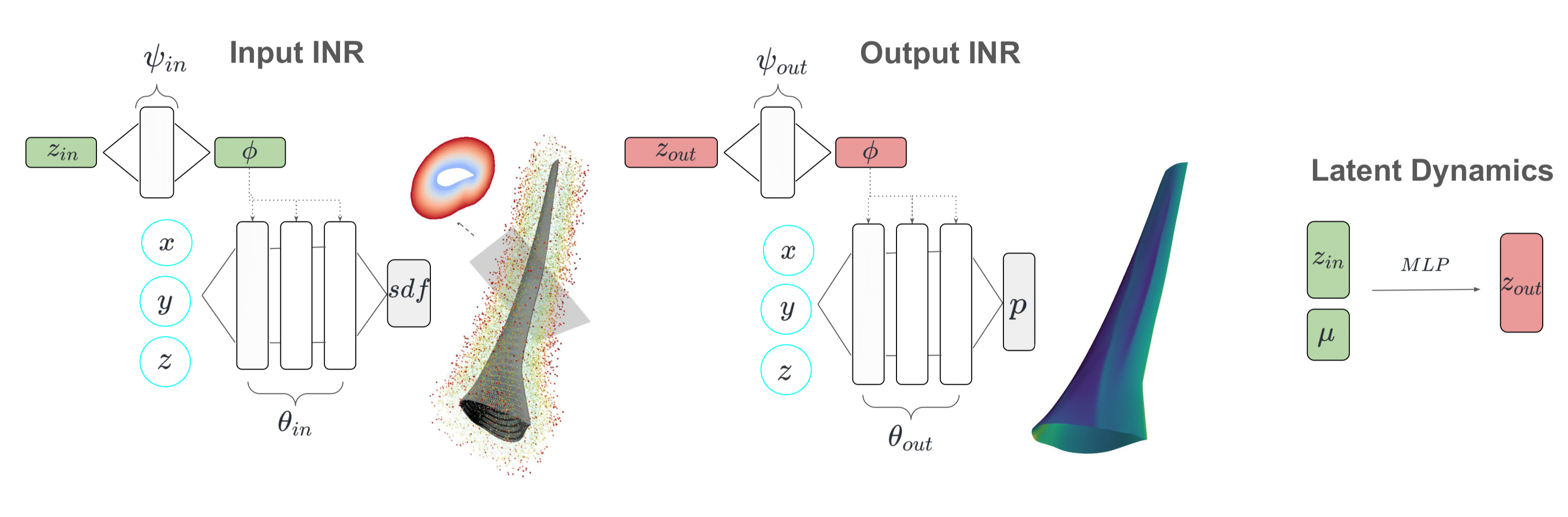}
        \label{sfig:epd-framework}
    \end{subfigure}%

    \vspace{2pt} % Adds vertical space between the figures

    \begin{subfigure}[b]{\textwidth}
        \centering
        \includegraphics[width=0.55\linewidth]{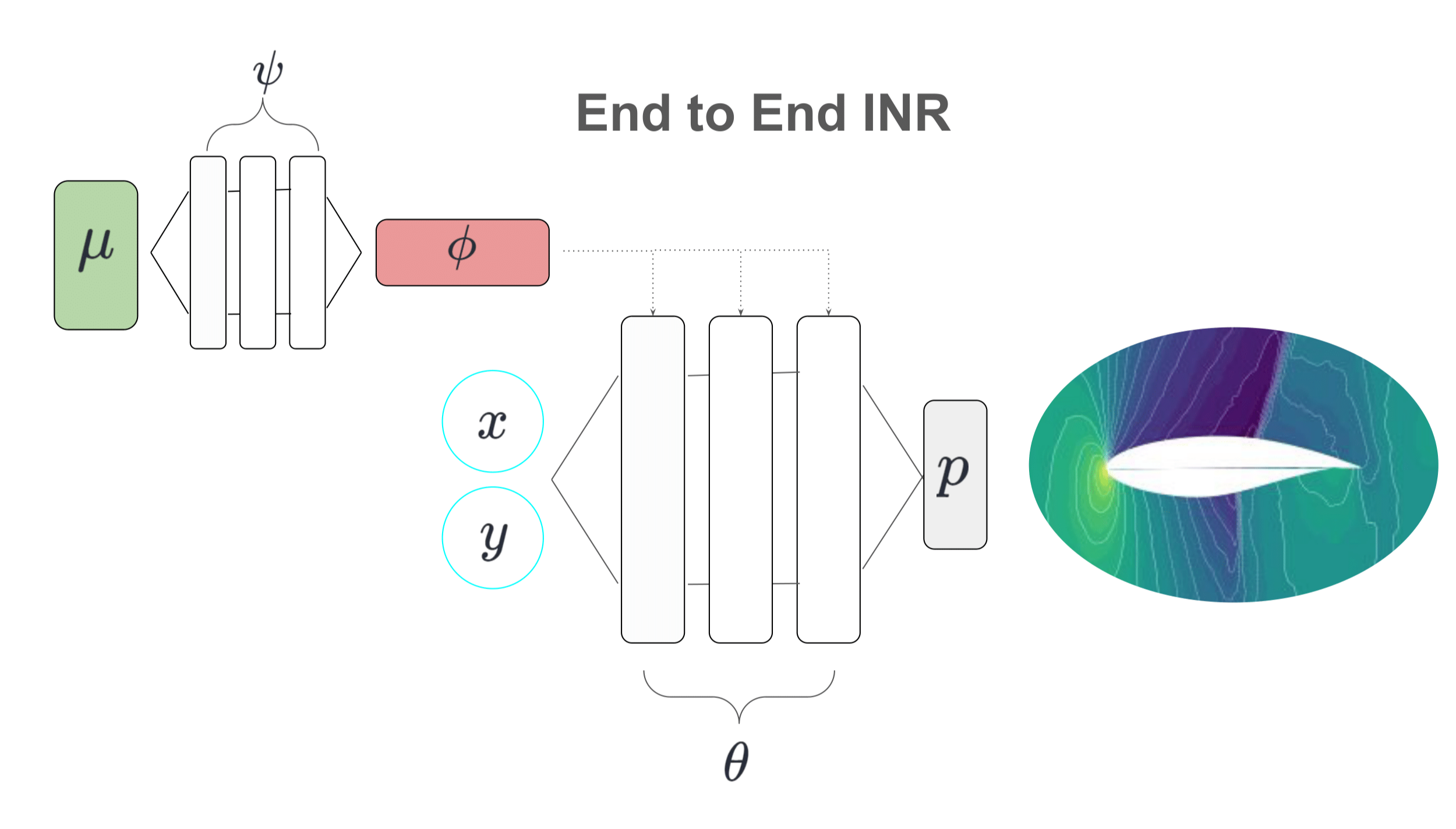}
        \label{sfig:endtoend-framework}
    \end{subfigure}%
    \caption{\textbf{Top:} The geometry is encoded by the input INR, where the spatial coordinates \(x, y, z\) are processed by a modulated neural network parameterized by \(\theta_{in}\) to produce a Signed Distance Function (SDF). The output INR, parameterized by \(\theta_{out}\), learns the mapping between the spatial coordinates and the output physical fields \(p\), for each training sample through modulation. The modulation vectors $\phi$ are obtained from the (input or output) latent codes $z$ through a hypernetwork parameterized by $\psi$.
   The latent map, where \(z_{in}\) and parameters \(\mu\) is transformed into \(z_{out}\), is approximated via a Multilayer Perceptron (MLP).
 \textbf{Bottom:} In this setup, the input parameters \(\mu\) are directly fed into a hypernetwork $h_{\psi}$ to produce the modulations $\phi$. This modulation, along with the spatial coordinates, is then processed by a neural network parameterized by \(\theta_{out}\) to output the physical fields \(p\).
  }
    \label{fig:frameworks}
\end{figure}

Figure~\ref{fig:frameworks} illustrates the components of the Implicit end-to-end framework.

\paragraph{Encode-Process-Decode framework}
This framework, proposed by Dupont et al. \cite{dupont2022data} and specialized by Serrano et al. \cite{serrano2024operator} for mesh based simulations, splits the learning process into two main steps: firstly learning compact representation of input and output field quantities ($z_{in}$ and $z_{out}$ in Fig. \ref{fig:frameworks}), and secondly learning the mapping between the two latent spaces ($z_{out}=MLP(z_{in},\mu$) conditioned by the input parameters.

The encoding step is analogous for both input and output variables (e.g. for the geometry and the pressure fields). We aim to model the geometry through the Signed Distance Function.
During the encoding step, for both input and output variables,  latent codes $z$, global networks $\theta$, and hyper-network parameters  $\psi$ are jointly optimized using a second-order Meta-Learning approach based on the CAVIA algorithm \cite{zintgraf2019fast}. The learning task can in fact be split into a sample-specific regression problem: to find the optimal $z$ specific to each distinct sample, and a global regression to learn the optimal values of the global shared networks parameters ($\theta,\psi$) across the whole dataset samples. A detailed description of the training algorithm can be found in Algorithm \ref{alg:CAVIA} and in the provided references.
As observed by Serrano \cite{serrano2024operator}, this strategy offers several advantages over classical approaches for training conditional Neural Fields: training history is stabilized and the risk of overfitting is reduced, as the latent parameters are re-initialized to zero at each epoch and learned in few gradient steps inside the inner loop, thus acting as a regularization term on the optimized latent vectors. Moreover, this speeds up remarkably the inference time, as only few optimization steps are required to learn representations of the input quantities compared to a classical approach as a larger amount of optimization steps would be required to obtain the latent input representations.
Encoding the input and output variables allows to obtain input-output latent vector pairs:
\begin{equation}
    \mathcal{D}^{\textit{enc}} = \{(z^{in}_i, \mu_i), z^{out}_i\}_{i=1}^{M} \quad z_{in}\in \mathbb{R}^{d_{in}}, z_{out}\in \mathbb{R}^{d_{out}}.
\end{equation}

The regression step entails learning a mapping in the compressed space of latent input vector ($z_{in}$), input parameters ($\mu$), and output latent vectors ($z_{out}$). For this task, we employ train a simple residual MLP $p_{\delta}$, with SiLu activations, by minimizing the latent loss:
\begin{equation}
    \min_{ \mathbf{\delta}} \sum_{i}^N [p_{\delta}(z^{in}_i, \mu_i) - z^{out}_i]^2 
\end{equation}

Figure~\ref{fig:frameworks} illustrates the architecture of the Encode-Process-Decode framework. \\

\begin{figure}[h]
  \centering
  \setlength{\unitlength}{1cm}
  \includegraphics[width=0.7\linewidth, trim=15cm 8cm 10cm 5cm, clip]{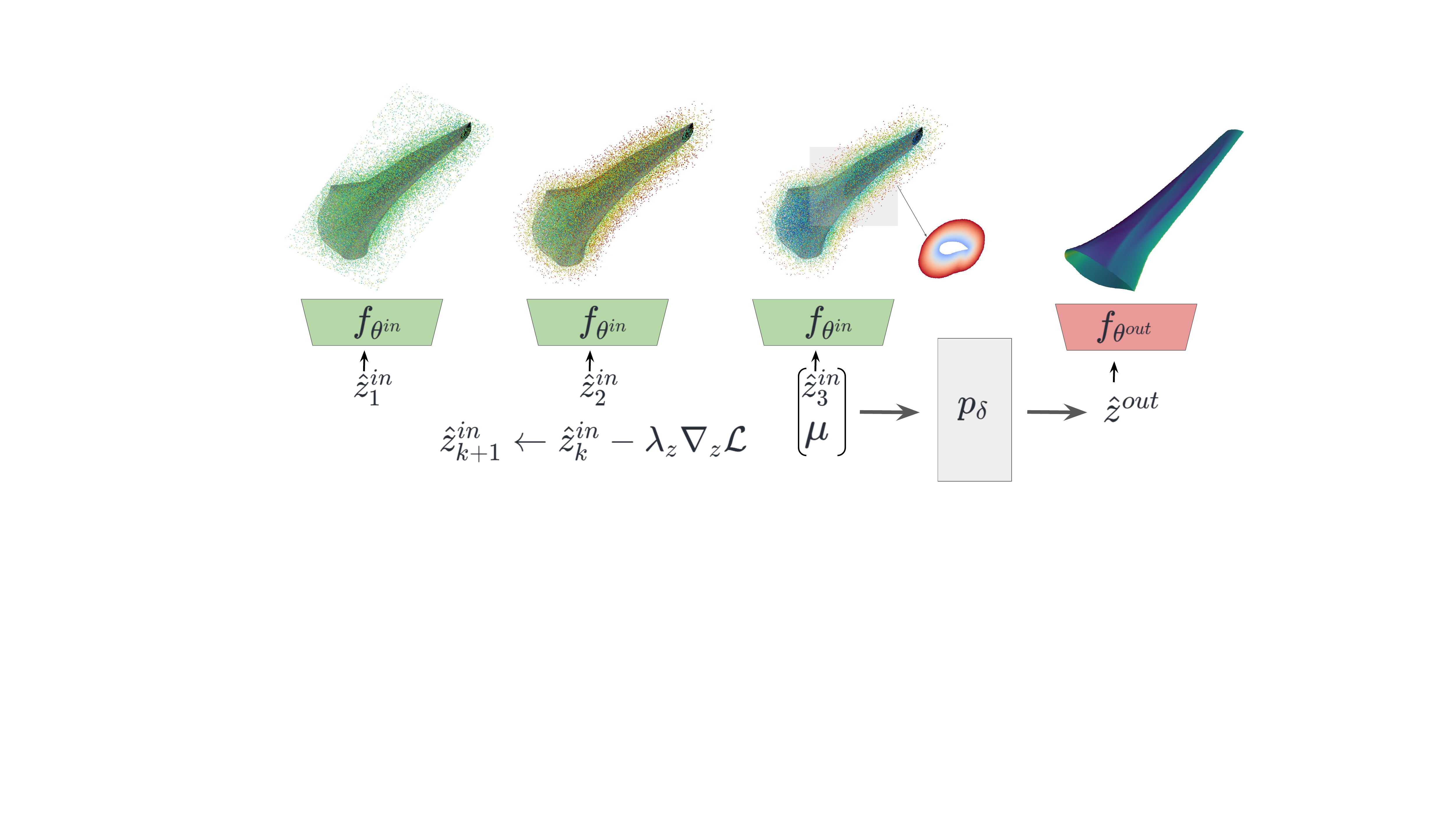}
  \vspace{-2.5cm}
  \caption{At inference time, few gradient steps are needed to optimize the input shape descriptors $\hat{z}^{in}$ to produce good reconstructions of the Signed Distance Function. The processor network $p_{\delta}$ maps the input latent and the parameters $\mu$ to the output $\hat{z}_{out}$, which allows decoding the output fields everywhere in the spatial domain.}
  \label{fig:inference}
\end{figure}

At \textbf{inference}, predictions for an unseen shape and flight parameters can be performed using the pretrained Neural Fields and processor network. 
In particular, the input latent code for an unseen shape can be inferred by minimizing the loss function ($\ell$) between the reconstructed input function (Signed Distance) and the target values:
\begin{equation}
    \Hat{z}^{in}_{test} = argmin_{z} \sum_{\mathbf{x} \in \mathcal{M}_{test}} \ell\left( f_{\mathbf{\theta}^{in}}(\mathbf{x}; h_{\mathbf{\psi}^{in}}(z)), u_{test}(\mathbf{x}) \right)
\end{equation}
practically, this corresponds to the just performing the inner loop described in Algorithm \ref{alg:CAVIA}, while keeping the global network parameters fixed.
Once the geometry descriptor is obtained, it can be used as an input to the processor network, together with the input parameters, in order to obtain the latent representation of the output variable:
\begin{equation}
    \Hat{z}^{out}_{test} =p_{\psi}(z^{in}_{test}, \mu_{test}).
\end{equation}
Finally, the output fields can be decoded on the spatial domain, by querying the pretrained output Neural Field:
\begin{equation}
    \Hat{u}_{test} =f_{\mathbf{\theta}^{out}}(\mathbf{x}; h_{\mathbf{\psi}^{out}}(\Hat{z}^{out}_{test}))
\end{equation}
The inference steps are also sketched in Fig. \ref{fig:inference} for a sample 3D wing shape.

\begin{algorithm}
\caption{Training Encoder Networks with CAVIA}
\label{alg:CAVIA}
\begin{algorithmic}[1]
\State \textbf{Input:} Dataset $\mathcal{D}$, epochs $E$, inner loops $K$, batch size $B$, learning rates $\lambda_{\theta}$, $\lambda_{\psi}$,$\lambda_{z}$
\State \textbf{Output:} Optimized parameters $\theta$, $\psi$
\For{epoch $e = 1, \ldots, E$}
    \State Sample $\mathcal{B}$ from $\mathcal{D}$
    \State Initialize context parameters $z_i = 0$ 
    \For{$i \in i_{\mathcal{B}}$ and $k$ in $1,..,K$} \Comment{Inner loop: Optimize local parameters}
        \State Compute sample loss $\mathcal{L}_i= \sum_{\mathbf{x} \in \mathcal{M}_{i}}\ell \left( f_{\mathbf{\theta}}(\mathbf{x}; h_{\mathbf{\psi}}(z_i)), u_{i} \right)$
        \State Update local parameters: $z_i \leftarrow z_i - \lambda_z \nabla_{z_i} \mathcal{L}_i$
    \EndFor
    \State Compute total loss for the batch: $\mathcal{L}_{\mathcal{B}} = \sum_{i \in i_{\mathcal{B}}} \mathcal{L}_i$
    \State Update global parameters: $\theta \leftarrow \theta - \lambda_{\theta} \nabla_{\theta} \mathcal{L}_{\mathcal{B}}$, and  $\psi \leftarrow \psi - \lambda_{\psi} \nabla_{\psi} \mathcal{L}_{\mathcal{B}}$
\EndFor
\end{algorithmic}
\end{algorithm}

\section{Experiments}
In this section we demonstrate our methodology on two distinct test cases. Firstly, a 2D RANS dataset over a fixed airfoil geometry, and a wide range of Angles of Attack and Mach numbers. Here the end-to-end framework is used and benchmarked against several baseline models.
The second test case, is a more challenging 3D wing dataset with shape and flight parameters variations: this allows us to showcase the encode-process-decode framework against a state of the art Graph Neural Network.
\subsection{Transonic Airfoil Dataset}
\paragraph{Dataset and Task Descriptiom}
The first test case considered to demonstrate the proposed methodology consists of Reynolds Averaged Navier Stokes (RANS) solution over a two dimensional transonic airfoil, namely the RAE2822. This geometry is well-documented in both numerical and experimental fluid mechanics research, ensuring a robust basis for validation.

The dataset encompasses a broad spectrum of aerodynamic conditions, with simulations carried out over a Reynolds number of 6.5 million, typical of cruise flight scenarios. The angle of attack and Mach number, critical parameters in aerodynamic studies, vary from 0° to 9° and 0.3 to 0.9, respectively, providing a dense sampling across both weakly compressible and transonic regimes. In total, the dataset consists in total of 1200 samples.

The RANS simulations, performed using the commercial software Fluent, leverage the Spalart-Allmaras \cite{spalart1992one} turbulence model and a second-order upwind discretization scheme on a hybrid mesh structure. The same hybrid CFD mesh,  consisting of 27499 cells, is used for all samples in the dataset, with a refined structured block close to the airfoil in order to accurately capture the boundary layer near the airfoil and an unstructured blocks, further away from the airfoil surface. Such detail ensures the fidelity and granularity required for high-quality machine learning training datasets. Each RANS computation takes approximately 30 minutes to converge to the steady-state solutions, utilizing an Intel Core i7-10850H processor with 6 cores. More details about the dataset can be found in the original paper introducing the dataset \cite{catalani2023comparative}.

The machine learning task consists in learning a surrogate model capable of predicting the output pressure field over the entire mesh (on and off the airfoil surface) for any combination of Mach number $M$ and angle of attack $\alpha$, given a training set of samples. We limit the discussion to one output quantity, as it is generally the most relevant variable for aerodynamics analysis and force estimation. It is possible, through the presented model, to predict additional quantities, such as the velocity components, using the same shared Neural Network: this is presented in the Appendix for the sake of completeness. We also note that, increasing the number of desired outputs, does not significantly change the memory requirements of the method. This is not the case for GNN based surrogates trained on large meshes, as storing the additional outputs on the full graph can become a limitation when training on memory constrained GPUs.

\subsubsection{Experimental Setting and Baselines}

Experiments are conducted using a 70/10/20 train/validation/test dataset split for all models. We compare the proposed INR end-to-end model with a simple MLP baseline and a state of the art Graph UNET Architecture. A brief explanation of the architectural choices for the proposed method and the baselines is provide in the following.

The INR end-to-end model is implemented as 5-Layer MLP with Multiscale Feature Encoding, ReLu activations and Shift Modulation, as outlined in Section \ref{sec:methodology}. For the input encoding, we sample Fourier Features with two distinct normal distributions with standard deviations $\sigma_1=1,\sigma_2=5$, using in total 128 components, aiming to cover the optimal frequency spectrum to fit the dataset . We observed that using larger values of $\sigma$ introduced more noisy reconstructions.
The hypernetwork is implememnted as  3-Layer MLP, processing the input parameters $\mathbf{\mu}=[M, \alpha]$ into latent modulations of dimension 128.
The model is trained on a downsampled version of the original mesh, as 5000 nodes for each samples are randomly selected for training. As it is shown in the following paragraphs, this allows to drastically reduce the training time, while keeping approximately invariate the test error at full resolution.
A more detailed overview of the hyperparamters and implementation details for the proposed method and the baselines is provided in the Appendix. 

\paragraph{Vanilla MLP}
A standard Multi-Layer-Perceptron architecture serves as a simple baseline, consisting of a 5-layer MLP with residual connections. The intermediate layer width is the same as the INR model, in order to keep a similar number of parameters and ensure a fair comparison.  For this baseline, the input parameters $\mathbf{\mu}$ are directly fed into the network along with the spatial coordinates, making it essentially a basic INR without latent conditioning and input encoding. This baseline helps to isolate and evaluate the specific impact of input encoding and latent modulation on model performance. 

\paragraph{Graph UNET}
Abbreviated GUNET, is a multiscale message passing graph convolutional neural network introduced in 2019 \cite{gao2019graph} that extends the popular UNET architecture \cite{ronneberger2015u} to handle non unfiform grids. Graph UNets are designed to handle graph-structured data by performing hierarchical pooling operations, which reduce the size of the graph at each level, and increase the receptive field of the convolutions, thereby capturing multi-scale features. We use an encode-process-decode architecture, where the encoder and decoder are implemented as a 4-layer MLP. The encoded features are passed through a SAGE Convolutional Layer \cite{hamilton2017inductive} followed by a pooling layer that uses TopKPooling \cite{cangea2018towards}. Each pooling layers, halves the number of nodes, and doubles the number of features. A multiscale GNN is particularly suited for this dataset, due to the moderately large mesh size, which requires more than local message-passing as in mono-scale architecture. However, we observe that it is possible (and beneficial), to train the model on a downsampled version of the orginal mesh and then to perform directly inference on the full mesh. This is not surprising, and was obsrved in previous work on MultiScale GNNs \cite{fortunato2022multiscale}. For this reason, the best performance of the model, is obtained when training on a mesh of 5000 nodes, and using 2 levels of Pooling. It must be observed that the mesh is dynamically downsampled at each epoch, in order to maximize variance and improve generalization error. This approach is analogous to the training strategy used in \cite{hines2023graph}, where a MeshGraphNet is trained on a large number of nodes representing a wing-shape.
A multiscale approach is necessary for this model because message passing on large graphs requires pooling to manage computational complexity and memory usage effectively, making it feasible to process large-scale graph data. The trained model is directly used for inference at full resolution at test time. In Section \ref{sec:rae_results} we provide quantitative results in the model performance, highlighting the effect of training resolution and number of pooling layers.

\paragraph{Evaluation Metrics} In our study, the performance of various models designed for predicting transonic airfoil aerodynamics is quantitatively assessed using three primary metrics: Mean Squared Error (MSE), training time, and inference time. These metrics provide a robust framework for comparing the efficiency and effectiveness of each modeling approach under consideration.

Training time refers to the total duration required to train a model on the specified dataset. Depending on the model type, it is specified into the different stages of the learning process. The experiments were conducted using a single NVIDIA's A100 GPU.

Inference time is defined as the average time required to perform a single sample prediction. When possible within the computational resources, batched inference can be performed, with additional speedup. In order to ensure fairness in comparison between the different baselines, we do not perform batching over the sample dimension. This metric is crucial for applications where real-time predictions are necessary, and it can vary drastically between different methodologies.

\subsubsection{Results}
\label{sec:rae_results}
The performance of the models on the transonic airfoil dataset is summarized in Table \ref{tab:comparison_rae}. 

\begin{figure}
  \centering
  \includegraphics[width=1\linewidth]{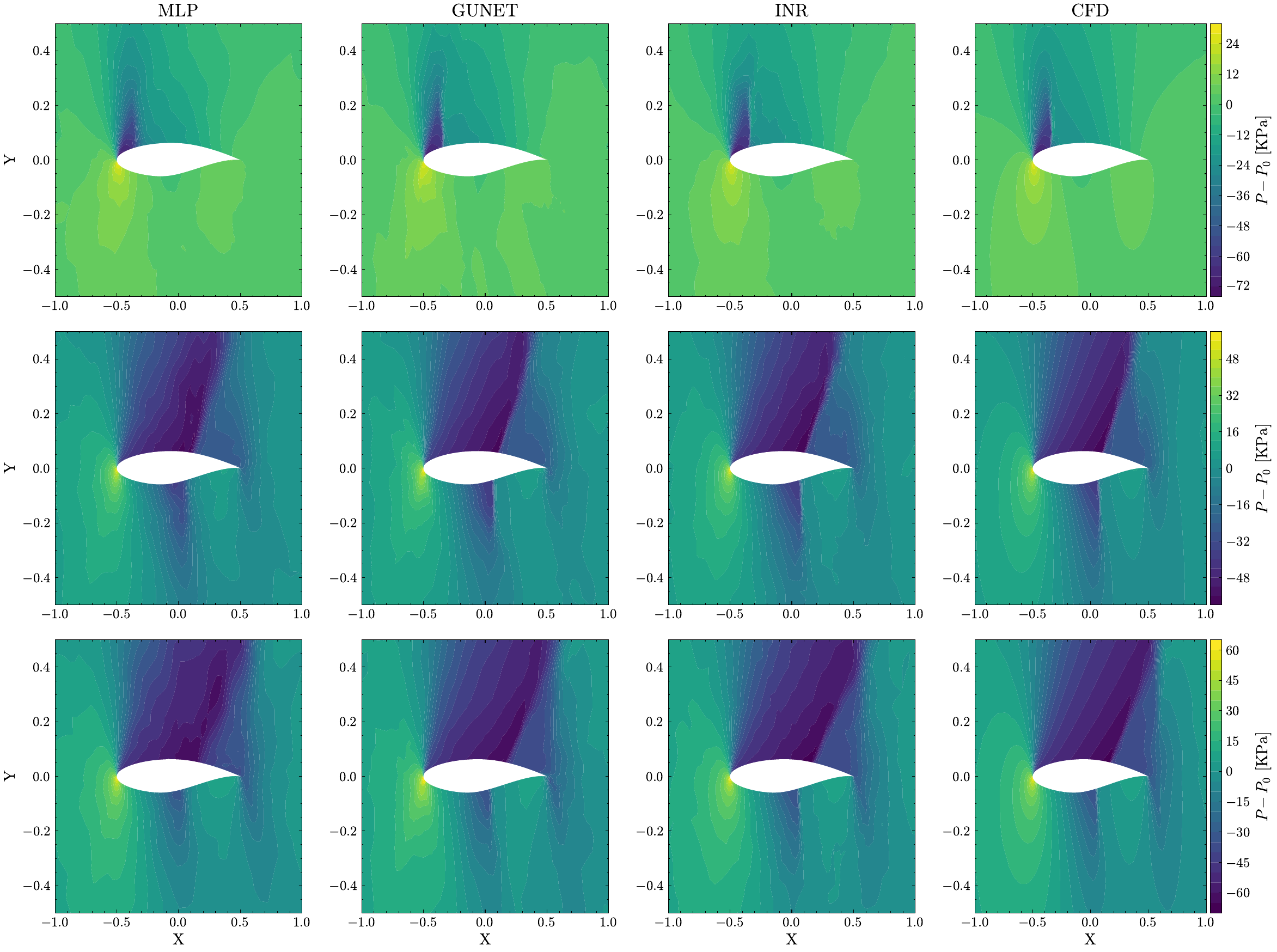}
  \caption{Pressure field distribution prediction with different surrogate models and comparison with CFD. \textbf{Top:} Angle of Attack 6.2 [deg], Mach Number 0.60. \textbf{Center:} Angle of Attack 4.94 [deg], Mach Number 0.84. \textbf{Bottom:} Angle of Attack 7.2 [deg], Mach Number 0.85}
  \label{fig:contour_rae}
\end{figure}

\begin{figure}
  \centering
  \includegraphics[width=1\linewidth]{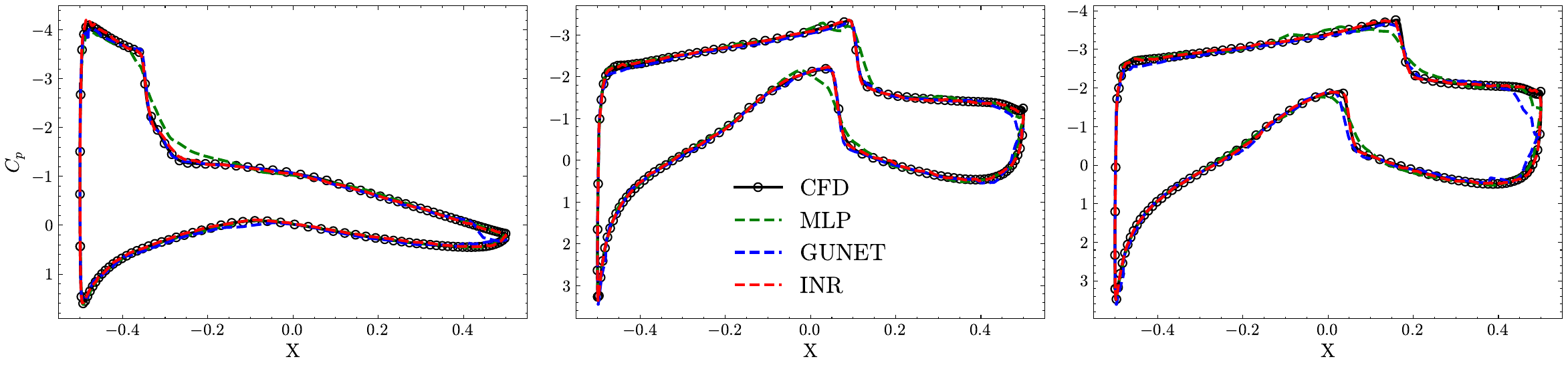}
  \caption{Surface Pressure Coefficient distribution prediction with different surrogate models and comparison with CFD. \textbf{Left:} Angle of Attack [6.2 deg], Mach Number [0.60]. \textbf{Center:} Angle of Attack [4.94 deg], Mach Number [0.84]. \textbf{Right:} Angle of Attack [7.2 deg], Mach Number [0.85]}
  \label{fig:cp_rae}
\end{figure}

In Figure \ref{fig:contour_rae}, we observe the performance of the models in capturing the pressure distribution in the spatial domain and on the airfoil surface. The simple MLP approach shows high accuracy away from the shock region, effectively capturing the pressure distribution in the weakly compressible regime. However, due to the spectral bias, and the absence of input encoding, the shock region is oversmoothed, failing to capture the sharp gradients accurately. This limitation is visible in the surface pressure distribution plot in Figure \ref{fig:cp_rae}, where the shock is noticeably smeared, and in the contour plots, as the isobaric lines are further apart in the vicinity of the compression wave. The trailing edge region, as well as the pressure peaks are well captured. Training and inference times for MLP-based approaches are relatively fast, with the vanilla MLP being especially quick due to the absence of input encoding and latent modulation.

The GUNET model, designed to handle graph-structured data with multi-scale features, performs well in the shock region prediction. The hierarchical pooling operations in GUNET enable it to capture multi-scale features, which helps in modeling the sharp gradients near the shock. However, its overall error is still higher than the MLP-based approaches, notably in the trailing edge region. Training time for GUNET is the longest due to the inclusion of pooling operations and multiple message passing steps. Despite this, GUNET offers significant speedup at inference time compared to computing high-fidelity solutions directly.

The INR model outperforms all other approaches, achieving the lowest MSE. The multi-layer MLP with multiscale feature encoding and shift modulation enables the INR to effectively model the complex aerodynamic phenomena present in the dataset. The incorporation of Gaussian feature encoding and latent modulations allows the INR to maintain high accuracy across different aerodynamic conditions, including the challenging transonic and shock regimes.
The speedup of all tested surrogate models, compared to the high-fidelity RANS computations is in the order of $10^5$, demonstrating the computational gain of using a surrogate model.

\begin{table}[h]
\centering
\caption{Performance comparison of different models in terms of error (MSE), training time (s), and inference time (ms).}
\label{tab:comparison_rae}
\begin{tabular}{@{}lccc@{}}
\toprule
\textbf{Model}      & \textbf{MSE} & \textbf{Training Time (s)} & \textbf{Inference Time (ms)} \\ \midrule
INR          &   0.0020    &     9497   &       4              \\
Vanilla MLP  &   0.0057    &     5749   &       2              \\
GUNET        &   0.0074    &     17100  &       17                  \\ \bottomrule
\end{tabular}
\end{table}

\paragraph{Discretization Invariance Analysis}

 %Typical deep Learning based models based on Convolutional Neural Networks and Transformers expect training data defined on regularly discretized domains (e.g. cartesian grid for CNNs), with the remarkable limitation of not being able to generalize across different type and levels of discretization. 
 A large class of Deep Learning based reduced order models can only predict test signals at the same resolution of the training signals, and are often coupled with interpolation routines to extract the value of the underlying signals at finer resolutions: this process can perform poorly, as subgrid dynamics is filtered out \cite{azizzadenesheli2024neural}.
 This consideration includes method based on Convolutional Neural Network,transformer based architectures and modal decomposition techniques. While it is true that the effect of subgrid dynamics on larger scales can be modeled by those architectures, inferring the values of the predicted signals at sub-pixel locations is problematic.
 Crucially, fluid flows are modeled by PDEs on continuous domains, and numerical methods seeks for the best approximation of the continuous operator mapping the input to the output function spaces on a discretized domain, without the limitation of mapping input and outputs of fixed resolution.
 These observations are of primary importance in the development of efficient surrogate models with powerful generalization properties: data used for training is in fact often defined on unstructured domains, with different connectivity and topology.
 Moreover, high resolution data might not always be abundant, preferring the use of mixed resolution training dataset. 
 An even more compelling consideration regarding surrogate models for CFD must take into account the scalability of the method to handle large mesh simulation, as the ones often used for real aeronautical applications. In these scenarios, surface meshes can be constituted by a number of nodes in the order of 1 Million. This resolution is required in order to accurately model strongly non-linear phenomenon in the boundary layer regions, or pressure discontinuities in the transonic regime. The amount of computational resources, namely memory footprint and training time, can represent the main limitation in training ML-based surrogates for this class of applications. 
 In practice, Graph Neural Networks have been scaled to handle full graph training on moderately large meshes by incorporating pooling techniques as part of multi-scale architecture designed to improve message passing over larger regions of the domain (in an analogous mannner of multiscale CNNs). The definition of pooling routines is however non-trivial and mesh dependent, making the resulting architectures only partially to generalize on graphs of similar topology. Additionally, multiscale architecture, while requiring a smaller number of message passing layers, do not solve the issue related to memory requirements (the graph at original resolution is loaded in GPU memory) and often incur in additional training time. Other techniques, based on graph mini-batching or domain decomposition have been employed in the literature to improve the computational efficiency of GNNs.
 Ideally, surrogate models that can generalize across different resolutions present the advantage of allowing training at much lower resolutions, while being able to predict accurately solutions at full resolution. Implicit Neural Representations are naturally formulated as coordinate based networks, without constraints on the signal discretization, and can perform zero shot superesolution at finer scales at test time.
 We quantify the trade-off between accuracy and efficiency, by performing an analysis on the approximate discretization invariance of the model, obtained on the RAE2822 dataset.
The study is devised as follows: multiple identical INR architecture (we use the end-to-end best performing architecture) are trained on different version of the datasets, obtained by randomly downsampling the mesh nodes at different resolution levels. The models are then used to predict the CFD solutions at increasing resolutions and the global metrics in terms of mean square error are summarized. In Table \ref{tab:discretization} the quantitative results are presented: the INR can be trained at much lower resolutions (on less than 2 \% of the full mesh), and perform super-resolution at test time with a contained error increase. The test error at full resolution converges to a minimum as the training resolution is increased, demonstrating the approximate discretization convergence property of Neural Operators. Optimal trade-off between test error and training time can be achieved by training the model with a 6 times lower resolution than the full mesh. It is important to stress that memory requirements do not scale with training resolution, as the INR are coordinate-based and do not require any sort of graph processing during training.
The GUNET performance in terms of test MSE does not improve when using finer training meshes. Using SAGE Convolutional Layer aids the message passing scheme to propagate information on the graph when training at higher resolutions, as the nodes neighborhoods radii are defined in the physical space. However, the accuracy results are biased towards the training resolution, with a test error that increases considerably when performing inference on the full mesh. It is interesting to note how the addition of a pooling level (compared to the 2-level baseline) degrades the performance of the network, while increasing computational cost.

\begin{table}[h]
\centering
\caption{Performance of models at different resolutions. The table shows the average mean squared error (MSE) and training time for INR and GUNET models at various training resolutions. GUNET levels refer to the levels of pooling used in the model.}
\label{tab:discretization}
\begin{tabular}{@{}lccccc@{}}
\toprule
\textbf{Model} & \textbf{Training Res} & \textbf{Test Res} & \textbf{Average MSE} & \textbf{Training Time (s)} \\ \midrule
\multirow{6}{*}{INR} 
                     & \multirow{2}{*}{500}   & 500  & 0.00294         & \multirow{2}{*}{2899} \\
                     &                        & Full & 0.00276  &      \\  \cmidrule(lr){2-5}
                     & \multirow{2}{*}{5000}  & 5000 &  0.00200       & \multirow{2}{*}{9497} \\
                     &                        & Full & 0.00202  &      \\ \cmidrule(lr){2-5}
                     & \multirow{1}{*}{Full}  & Full  &  0.00197       & 46985 \\
                    \midrule
\multirow{6}{*}{GUNET (2 levels)} 
                     & \multirow{2}{*}{500}   & 500  &   0.0059       & \multirow{2}{*}{6080} \\
                     &                        & Full &  0.0078  &      \\ \cmidrule(lr){2-5}
                     & \multirow{2}{*}{5000}  & 5000 &  0.0038         & \multirow{2}{*}{17100} \\
                     &                        & Full & 0.0074   &      \\ \cmidrule(lr){2-5}
                     & \multirow{1}{*}{Full}  & Full  &   0.0079      &  35580    \\
                     \midrule
\multirow{1}{*}{GUNET (3 levels)} 
                     & \multirow{1}{*}{Full}  & Full  &   0.0113      &  43320    \\
                    
\bottomrule
\end{tabular}
\end{table}

\subsection{XRF1 Wing Dataset}
\begin{figure}[h]
    \centering
    \begin{subfigure}[b]{0.5\textwidth}
        \centering
        \includegraphics[width=\linewidth]{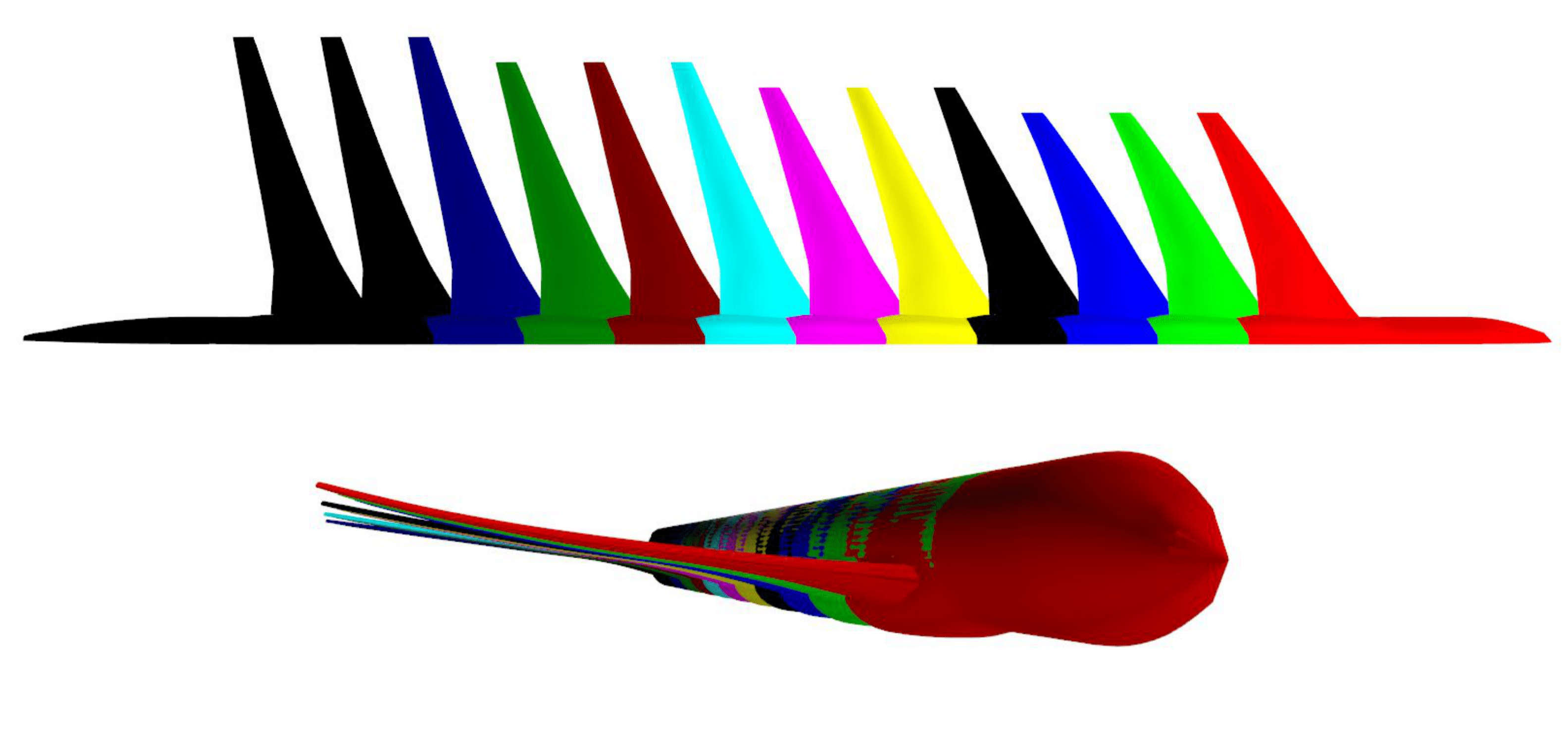}
    \end{subfigure}
    \hspace{-1em} % Adjust the space between the subfigures by changing the value
    \begin{subfigure}[b]{0.4\textwidth}
        \centering
        \includegraphics[width=\linewidth]{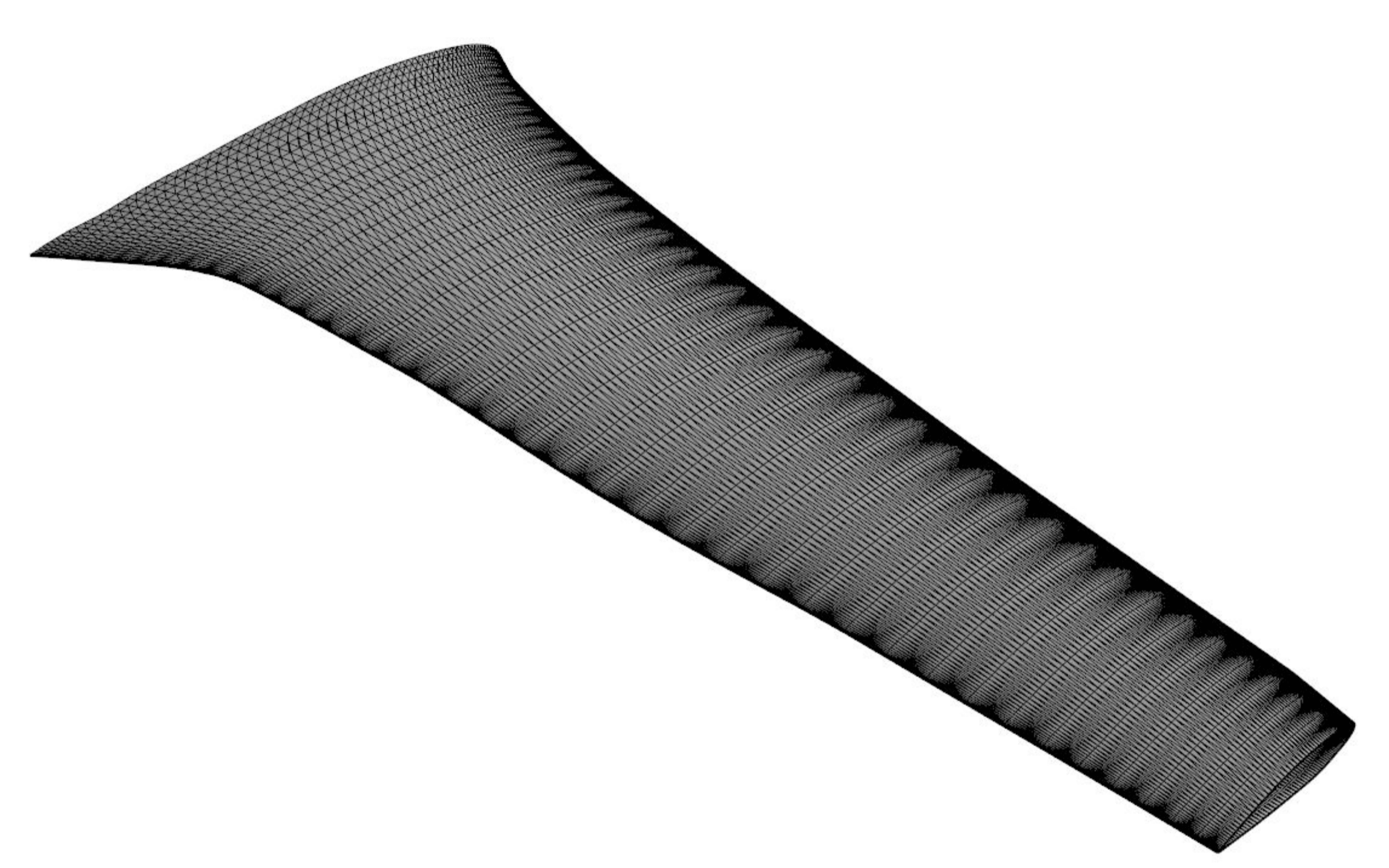}
    \end{subfigure}
    \caption{XRF Wing Dataset geometries.\textbf{Left:} Top and front view of 10 out of 120 different wing shapes. Span, shape and thickness are varied.\textbf{Right:} Sample computational mesh for a given shape.}
    \label{fig:xrf_dataset}
\end{figure}

\paragraph{Dataset Description}
The dataset comprises 8,640 surface pressure simulations over 120 wing shapes, and different operating conditions. The numerical solver is based on the BLWF method \cite{zhang2010evaluation} developed at the Central Aerohydrodynamic
Institute (TsAGI). This code assumes external inviscid flow coupled with a viscous boundary layer. For each computation, we only consider the pressure field on the wing, computed on a triangular mesh of 6,600 nodes. An example of the mesh for a specific shape is displayed in Figure \ref{fig:xrf_dataset}.
 Despite using a lower fidelity numerical solver, compared to typical 3D RANS computations on larger meshes, the relevant flow structures and pressure patterns are well captured, as a function of the wing geometries, ensuring the validity of the proposed surrogate on higher fidelity datasets.
Moreover, the dataset's diversity and complexity make it highly relevant for industrial aerospace applications. Testing generalization to unseen geometries and configurations in 3D is in fact crucial for a surrogate model to be effectively used within preliminary aerodynamic design and analysis of an aircraft.

The dataset spans 120 distinct wing configurations characterized by systematic variations of a reference wing, by variation of three shape parameters: Span $S$,Thickness $T$ and Dihedral Sweep $D_s$. In Figure \ref{fig:xrf_dataset} a subset of wing shapes used in this study is depicted.
For each configuration, operational conditions are defined with a simple Design of Experiment (DoE) approach, by changing the Mach number within typical cruise values $M \in [0.50,0.86]$, the Angle of attack $\alpha \in [-4^{\circ},8^{\circ}]$ and the Aileron Deflection Angle $\delta_{ail}\in [0^{\circ},-15^{\circ}]$. The Reynolds number is kept fixed at $Re=3\times 10^6$. The effect of the aileron deflection is modeled on the resulting pressure distribution, but not present in the input geometry and it is therefore kepts as a paramteric input for the surrogate model.
This provides a wide range of aerodynamic conditions, in the compressible and transonic regime, including control surface deflection effects.

The machine learning task consists of predicting the pressure fields on the surface mesh from the operational parameters $\mathbf{\mu}=[M,\alpha,\delta_{ail}]$ and the geometry information. It is important to note that here we assume non-parametric geometry definition: although the shape variations are defined by modification of the shape parameters, we decide to learn a representation of the geometry only from the mesh node position, connectivity and surface normals. This demonstrated the general capabilities of the proposed approach, since parametrization of CAD degigned geometries is not in general available. Learning concise representation of shapes is essential for downstream applications, such as shape optimization or shape generation.
\subsubsection{Experimental Setting and Baselines}
In this study, we compare the performance of the INR model, using the encode-process-decode framework, with a Graph Neural Network, specifically the MeshGraphNet, which is widely used for this type of task \cite{pfaff2020learning,hines2023graph}. 

\paragraph{Geometry Preprocessing and SDF} Training the input INR model requires a preprocessing step of the surface meshes to obtain the volumetric Signed Distance Function fields. We follow a similar process to the one applied to train the DeepSDF model of Park et al. \cite{park2019deepsdf}. In particular a bounding box is created around each mesh, prenormalized to the unit sphere.
A first random sampling, with uniform distribution, is performed inside the box ($10 \%$ of the total sampled points). The second sampling is performed by randomly selecting points on the mesh surface, and then adding random Gaussian noise with zero mean and pre-defined standard deviation. Two different values ($\sigma_{s}=0.005,0.0005$) are used, defining a progressively finer distribution of points near the surface. This is a fundamental step to capture the finer geometrical details of a specific wing. In total, around $N^{vol} = 60000$ points are sampled for each mesh.
Finally, the SDF is computed in each  of the sampled points, using the \texttt{point-cloud-utils} \cite{point-cloud-utils} library. A set of input coordinates and ouptut SDF values is obtained for each shape:
\begin{equation}
    \mathcal{D}^{SDF}_i = \{(\mathbf{x}_j, sdf_j)\}_{j=1}^{N^{vol}} \quad i=1,...,M
\end{equation}

\paragraph{Encode-Process-Decode INR}
The input INR is trained to learn the mapping between coordinates and sdf values, conditioned by latent vectors specific for each geometry $\{z^{in}_i\}_{i=1}^M$. The architecture is a 6-Layers MLP with a single scale Fourier Feature Encoding. Latent modulations are obtained from latent codes $z_{in}$ of dimension 64, with a single linear layer hypernetwork. In order to make training more efficient, we randomly select mini-batches of points $\mathcal{B} \in \mathcal{D}^{SDF}_i$ for each epoch.
The output INR maps the input node coordinates and the surface normals to the output pressure values. The architecture is a 6-Layer MultiScale FFN with $\sigma_1=1,\sigma_2=5$, single linear layer hypernetwork and latent codes  $z_{out}$of dimension 128.
A processor network is used to learn the mapping from input latent codes and the flight parameters to output latent codes. The flight parameters are stored in a 3-dimensional vector, containing respectively the Angle of Attack, the Mach Number, the Aileron Deflection. This regressor model is implemented as a 4-Block MLP with skip connections and SiLu activation. Training the processor is performed for a maximum 1000 epochs and stopped when validation loss has not improved for more than 200 epochs.

\paragraph{MeshGraphNet}
Abbreviated MGN, is employed as a baseline for this task, being widely used in literature for learning unsteady fluid dynamic \cite{pfaff2020learning}, and steady parametric solutions \cite{hines2023graph} on general meshes. 
MGN follows an encode-process-decode architecture without pooling layers. Given the manageable size of the triangular mesh (6,600 nodes), pooling layers are not required for effective message passing. 
The input features are specified on edges for relative node distance magnitude and sign, and on nodes for parameter values and normals.
Two distinct 4-Layer residual MLP networks implement the edge and node encoders, projecting the respective features to latent vectors (one latent vector for each node or edge) of dimension 128.

The processor block includes 12 consecutive Graph Convolutional Layers with multiple consecutive message passing layers with Edge Based Convolution. Across each layer, edges representations are updated a function (parameterized by an MLP) of their current state of the representations of the nodes connected by said edge, while the nodes representations are updated as a function of their current state and the sum of the edge representations converging to the node. The decoder projects the output of the processor block to the desired output features.  The specific architectural choices and hyperparameters are given in the appendix.

\subsubsection{Results}
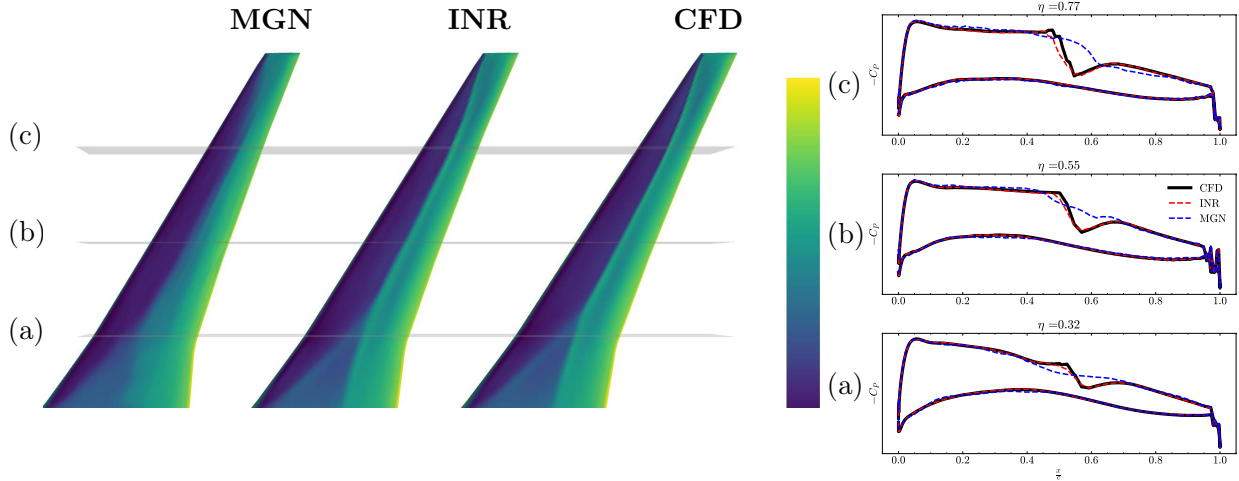
\begin{figure}[h!]
  \centering
  \begin{minipage}[b]{0.65\linewidth} % Adjust the width of the first subfigure
    \raisebox{1cm}{ % Adjust this value to move the subfigure up
      \centering
      \setlength{\unitlength}{1cm}
      \begin{picture}(0,0) % Create a picture environment with zero size
        \put(2.6,5.3){\makebox(0,0)[lt]{\textbf{MGN}}} % Adjust coordinates as needed
        \put(5.5,5.3){\makebox(0,0)[lt]{\textbf{INR}}} % Adjust coordinates as needed
        \put(8.5,5.3){\makebox(0,0)[lt]{\textbf{CFD}}} % Adjust coordinates as needed
        \put(-0.35,3.8){\makebox(0,0)[lt]{(c)}}
        \put(-0.35,2.5){\makebox(0,0)[lt]{(b)}}
        \put(-0.35,1.2){\makebox(0,0)[lt]{(a)}}
      \end{picture}
      \includegraphics[width=\linewidth, trim=23cm 19cm 1cm 10cm, clip]{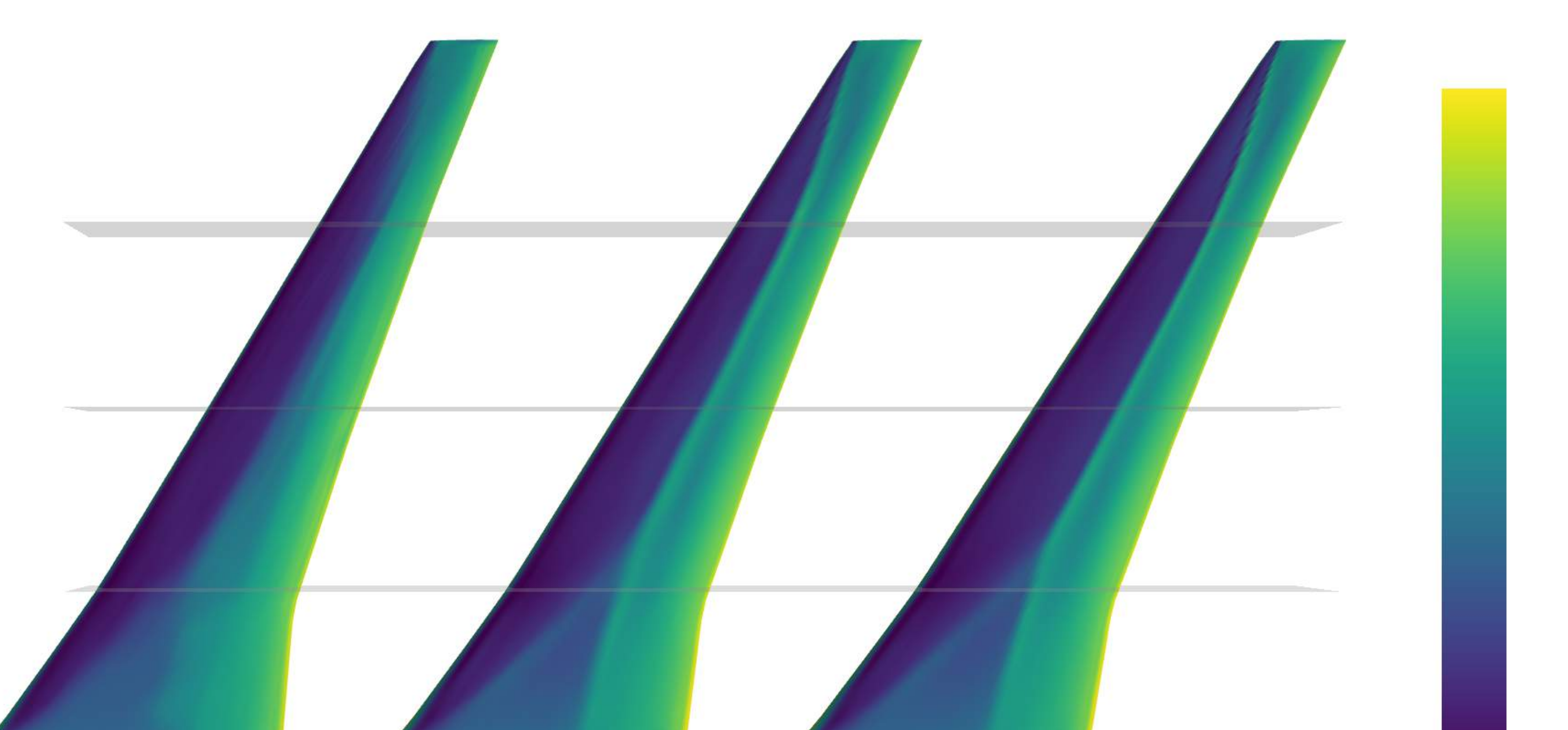}
    }
  \end{minipage}
  \hspace{0.02\linewidth} % Reduce the space between the two subfigures
  \begin{subfigure}[b]{0.3\linewidth}
    \centering
    \setlength{\unitlength}{1cm}
    \begin{picture}(0,0)
      \put(-3,-1){\makebox(0,0)[lt]{(c)}}
      \put(-3,-3.0){\makebox(0,0)[lt]{(b)}}
      \put(-3,-5){\makebox(0,0)[lt]{(a)}}
    \end{picture}
    \includegraphics[width=\linewidth, trim=0cm 0cm 0cm 0cm, clip]{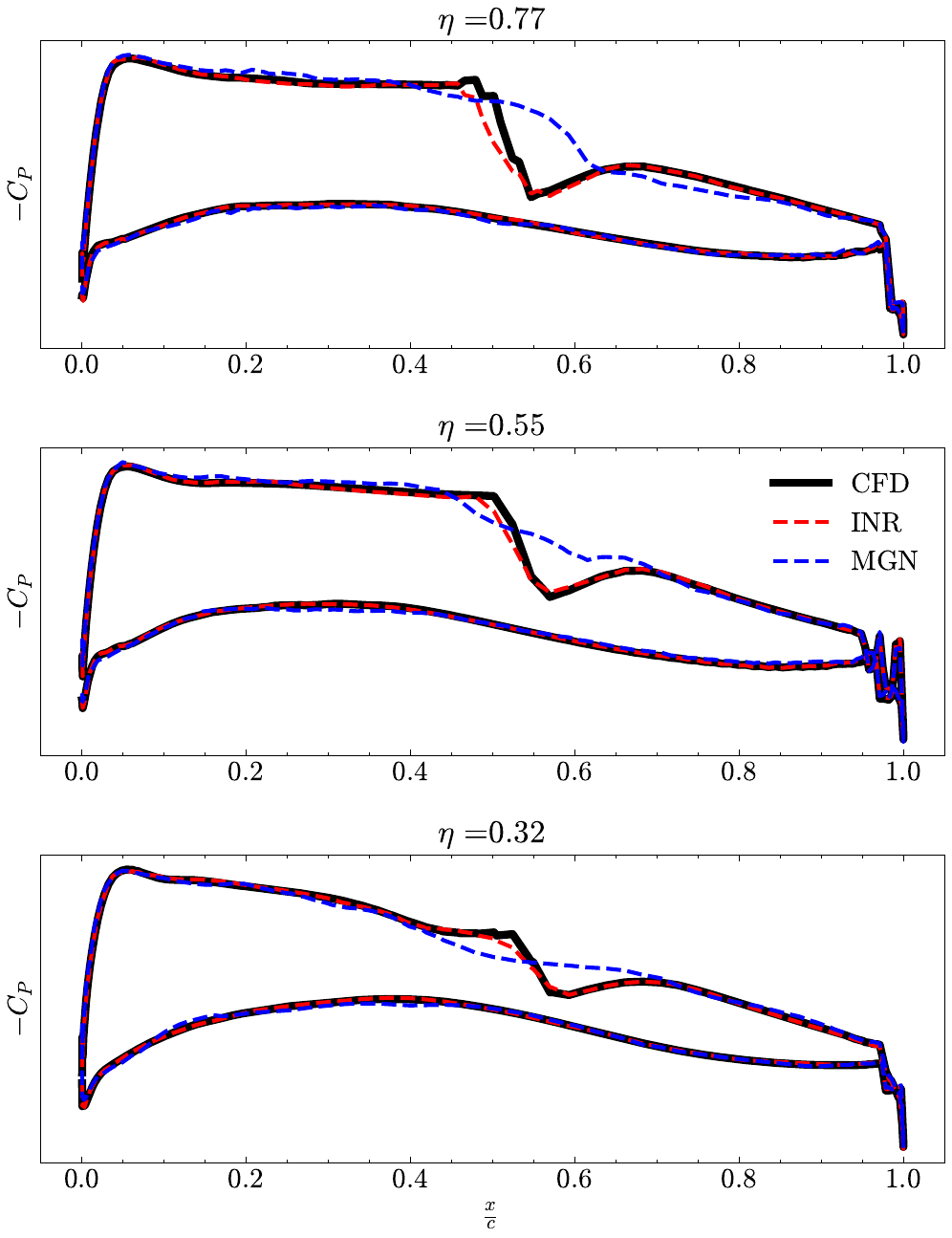}
  \end{subfigure}
  \caption{Comparison between MeshGraphNet, INR, and CFD predictions of the surface pressure distribution at $M=0.82$,$\alpha= 4 [deg]$,$\delta_{ail}=-15 [deg]$
  \textbf{Left: }Surface pressure distribution on the upper surface of the wing (top view). Cutting planes indicate the location of pressure distribution on the right. \textbf{Right: }Pressure coefficient cuts at three different locations $\eta=0.32,0.55,0.77$ along the wingspan.}
  \label{fig:contour_cp_xrf}
\end{figure}

The performance of the models on the XRF1 wing dataset is summarized in Table \ref{tab:model_performance_comparison}. The INR framework demonstrates far superior accuracy with the lowest MSE for predictions on unseen shapes. In Figure \ref{fig:contour_cp_xrf}, the surrogate models' predictions are displayed against the ground truth CFD data. The MGN method can capture the general trend of the pressure distribution, while failing to model the more complex aerodynamic features. The shock region prediction is the most challenging for both surrogate models. Nevertheless, the INR method is capable to predict the sharp gradient with good accuracy. Towards the leading edge, the accuracy of MGN degrades, as the pressure distribution shows strongly non-linear pattern due to the presence of strong vorticity. 

\begin{figure}[h!]
  \centering
  \begin{minipage}[b]{1\linewidth} % Adjust the width of the first subfigure
    \raisebox{0cm}{ % Adjust this value to move the subfigure up
      \centering
      \setlength{\unitlength}{1cm}
      \begin{picture}(0,0) % Create a picture environment with zero size
        \put(3,4.7){\makebox(0,0)[lt]{\textbf{MGN}}} % Adjust coordinates as needed
        \put(9,4.7){\makebox(0,0)[lt]{\textbf{INR}}} % Adjust coordinates as needed
        \put(14.,4.7){\makebox(0,0)[lt]{\textbf{CFD}}} % Adjust coordinates as needed
      \end{picture}
      \includegraphics[width=\linewidth, trim=2cm 7cm 0cm 4cm, clip]{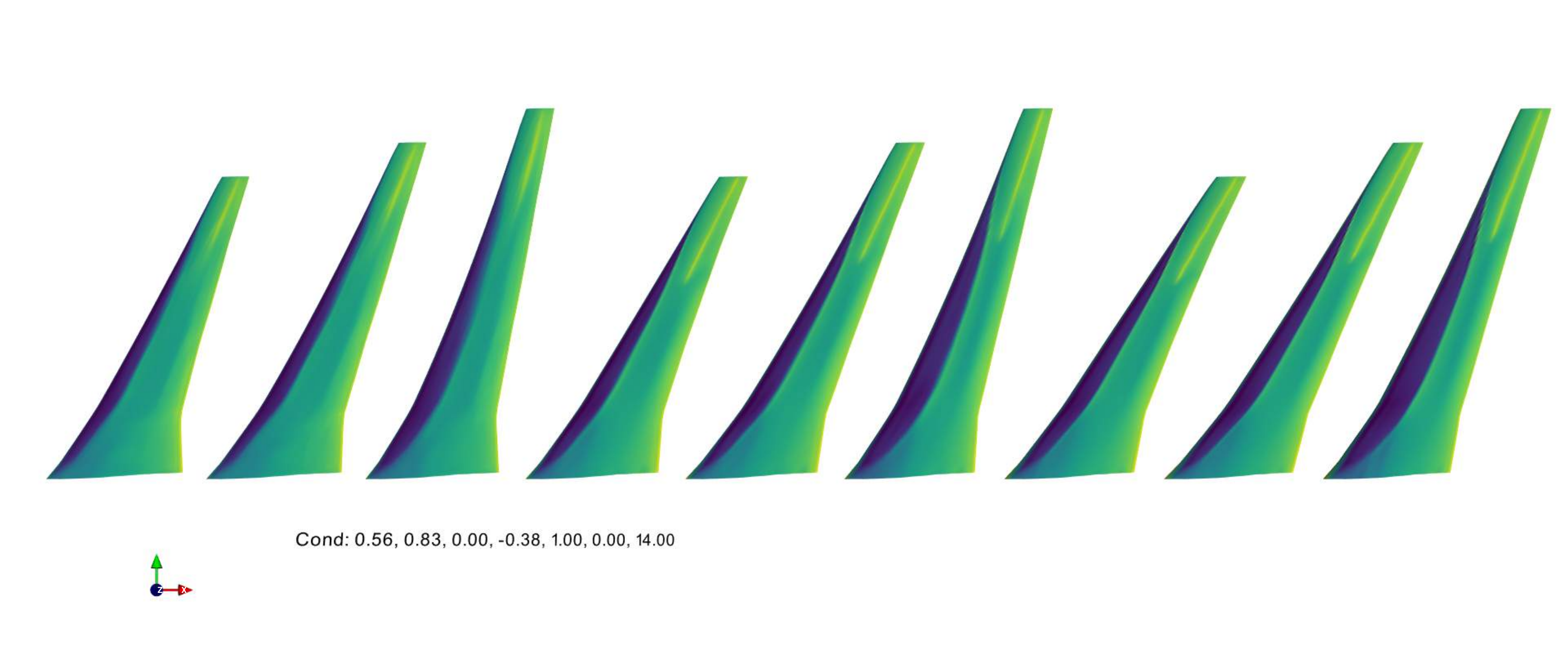}
    }
  \end{minipage}
  \vspace{0.2 cm} % Add space between the two images
  \begin{minipage}[b]{1\linewidth} % Adjust the width of the second subfigure
    \raisebox{0cm}{ % Adjust this value to move the subfigure up or down
      \centering
      \setlength{\unitlength}{1cm}
      \includegraphics[width=\linewidth, trim=2cm 7cm 0cm 4cm, clip]{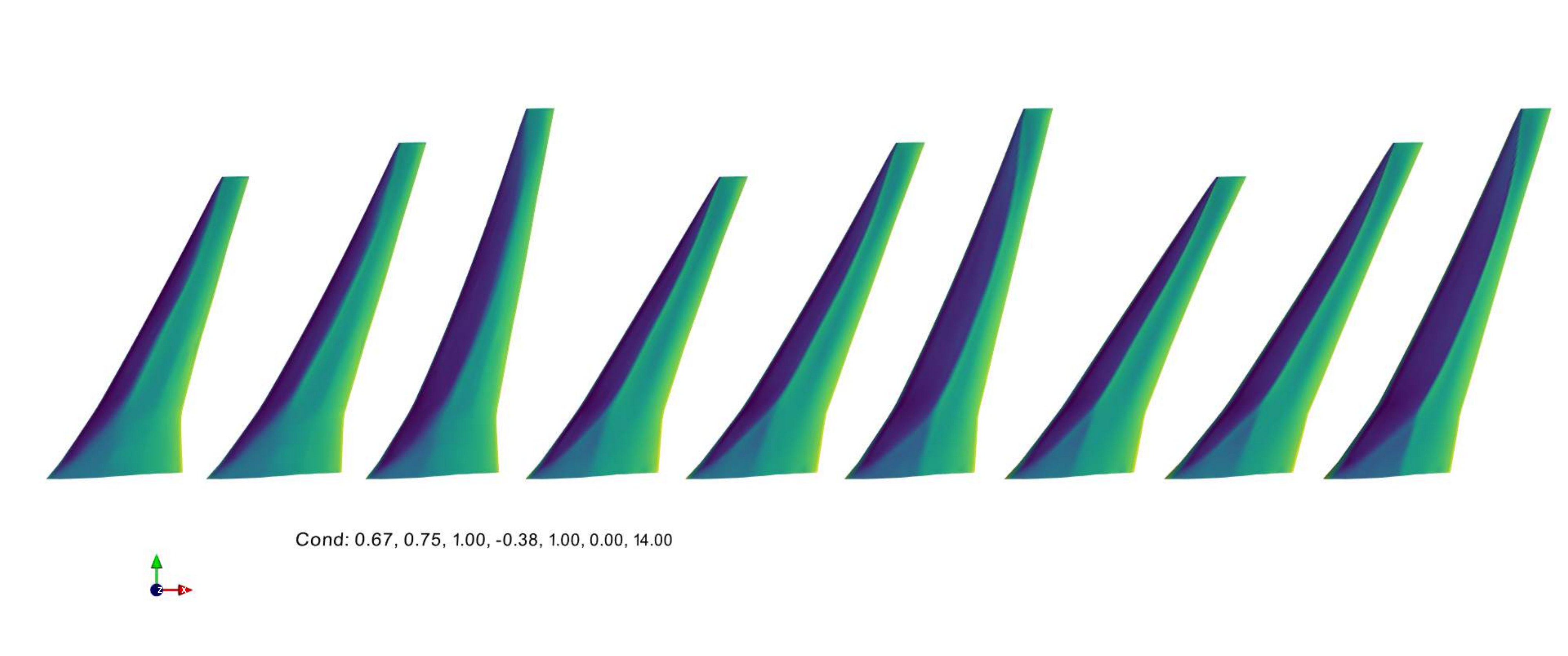} % Update the file path for the second image
    }
  \end{minipage}
  \caption{Pressure Distribution prediction on top surface, across different shapes at the same operating conditions. \textbf{Top: }$M=0.70$,$\alpha=8 [deg]$,$\delta_{ail}=0 [deg]$. \textbf{Bottom: }$M=0.80$,$\alpha=3 [deg]$,$\delta_{ail}=0 [deg]$.}
  \label{fig:contour_shapes}
\end{figure}
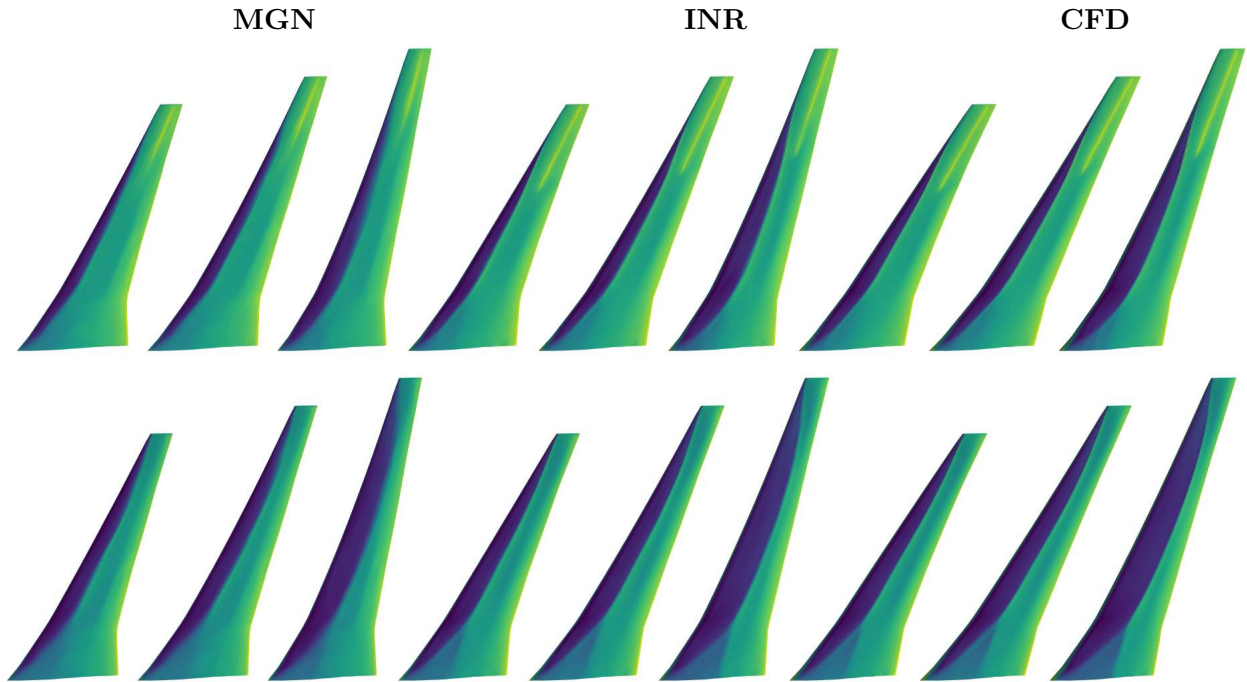

The INR method shows remarkable generalization capabilities to unseen shapes, demonstrating the efficacy of the geometry latent representations. This is visible in Figure \ref{fig:contour_shapes}, as the surrogate models' predictions on three different test shapes, at the same operating conditions, are presented against the CFD ground truth.

The training time for the INR encode-process-decode model is split into the three components: encoding the input (geometry), encoding the output (pressure fields), and the processor network. The input INR, which learns the Signed Distance Function (SDF) representation of the geometry, and the output INR, which maps coordinates to pressure values, both require significant training time. As the number of shapes is smaller than the number of total samples (for each shape a number of operating conditions are simulated), training the input encoder takes less time than the output encoder. The processor network, on the other hand, can be trained to fit the output latent codes using relatively limited computational resources. Similarly, the MeshGraphNet's training time is relatively high due to the extensive message passing and separate node and edge encoding.The inference time, for the proposed INR approach, is dominated by the geometry encoding step : as explained in the Section \ref{sec:methodology}, thanks to the meta-learning approach, the latent codes for a test geometry can be inferred in few gradient steps while freezing the main network parameters. It must be noted that this step is only required once per geometry. Thus, INR model can be used to decode the output fields for different operating conditions, on the same shape, without incurring in this computation. In a typical design scenario, in fact, the same shape design is tested over a variety of flight conditions relevant for the aircraft mission.

As a conclusion, these results highlight the effectiveness of the encode-process-decode framework in handling complex aerodynamic datasets with varying geometries and operating conditions.
\begin{table}[h]
\centering
\caption{Performance comparison of different models in terms of error (MSE), training time (s), and inference time (ms).}
\label{tab:model_performance_comparison}
\begin{tabular}{@{}lccc@{}}
\toprule
\textbf{Model}      & \textbf{MSE} & \textbf{Training Time (s)} & \textbf{Inference Time (ms)} \\ \midrule
\multirow{3}{*}{INR} & \multirow{3}{*}{0.008} & 28320 (INR Input) & \multirow{2}{*}{108 (Geometry)} \\
                     &                        & 73560 (INR Output)    &                             \\
                     &                        & 3318 (Processor)  & 9 (Output)                  \\ \cmidrule{1-4}
MGN                  & 0.035                 & 90720              & 17                          \\
\bottomrule
\end{tabular}
\end{table}

\begin{figure}[h]
  \centering
  \begin{minipage}[b]{0.5\linewidth} % Adjust the width of the first subfigure
    \raisebox{0.2cm}{ % Adjust this value to move the subfigure up
      \centering
      \setlength{\unitlength}{1cm}
      \includegraphics[width=\linewidth, trim=8cm 15cm 5cm 15cm, clip]{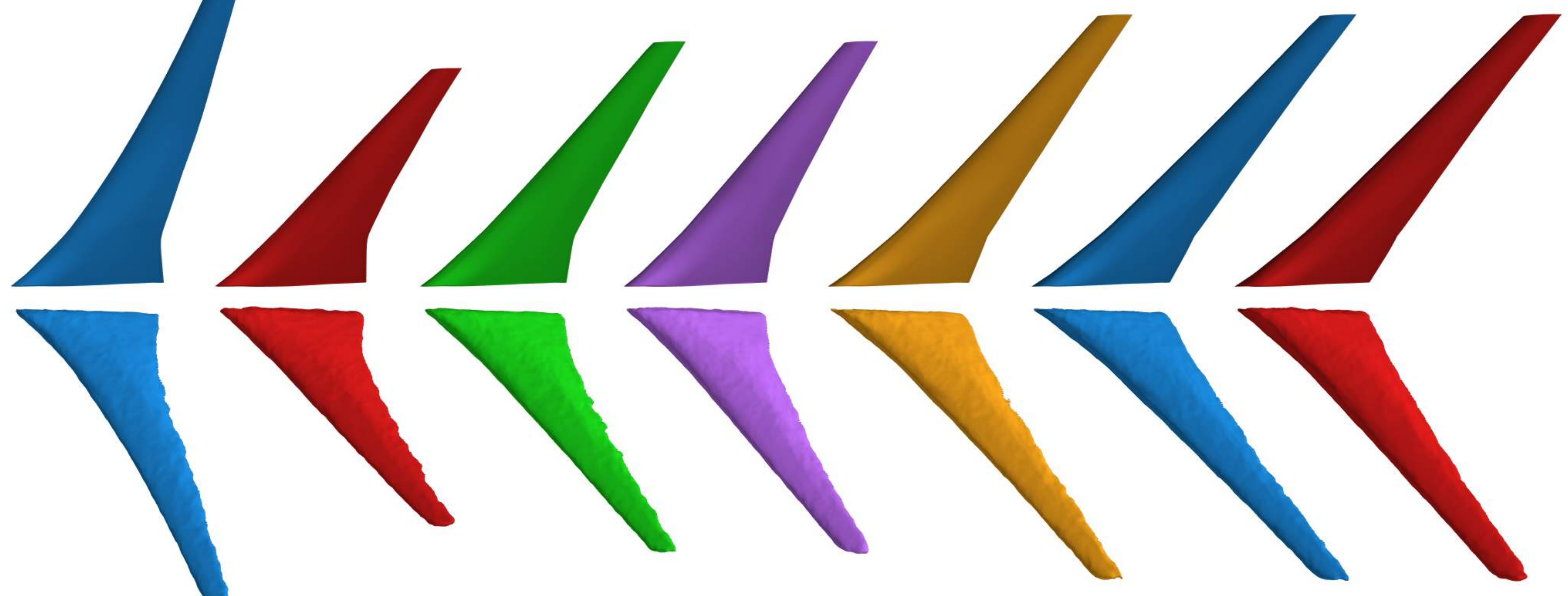}
    }
  \end{minipage}
  \hspace{0.01\linewidth} % Reduce the space between the two subfigures
  \begin{minipage}[b]{0.45\linewidth} % Adjust the width of the first subfigure
    \raisebox{0cm}{ % Adjust this value to move the subfigure up
      \centering
      \setlength{\unitlength}{1cm}
      \includegraphics[width=\linewidth, trim=5cm 5cm 1cm 5cm, clip]{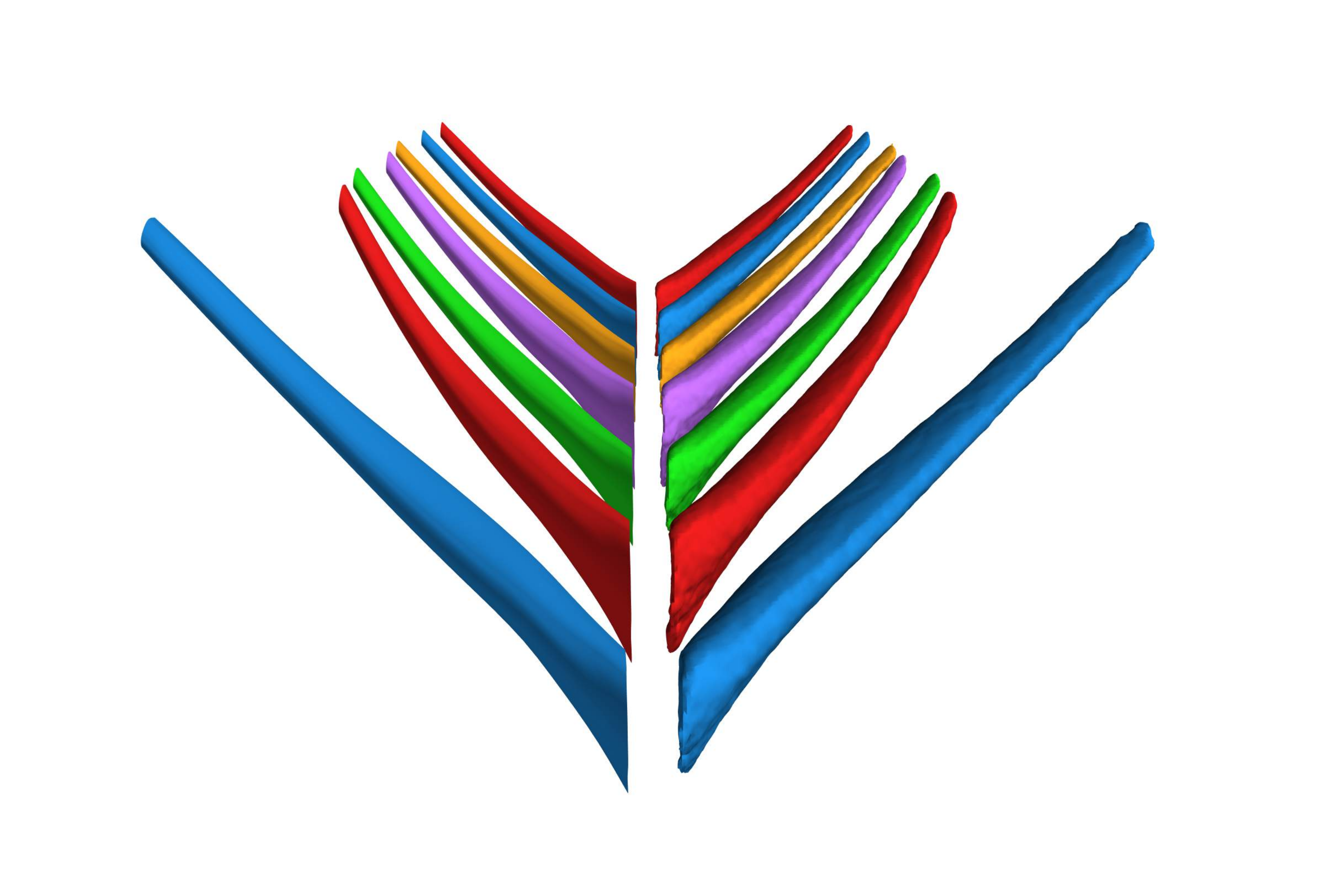}
    }
  \end{minipage}
  \caption{ \textbf{Left}: Ground Thruth (top) vs reconstructed (bottom) wing shapes, top view. \textbf{Right:} Ground Thruth (left) vs reconstructed (right) wing shapes, front view.}
  \label{fig:shape_reconstruction}
\end{figure}

\paragraph{Geometry encoding and reconstruction} A key factor in the effectiveness of the INR methodology is the possibility to effectively encode the geometrical information inside the input latent codes, by fitting a modulated Neural Field on the Signed Distance Function scalar field. In order to capture the shape variations in the dataset, a good level of accuracy is required, so that the latent representations unambiguously carry the specific geometrical features of the encoded objects. However, at test time, inferring the latent code for an unseen shape should not be a computationally demanding task, especially for surrogate modeling applications, where the main focus is placed on speedup and accuracy on the predicted physical fields rather than extremely detailed shape reconstruction. For this reason, the employed training strategy detailed in Algorithm \ref{alg:CAVIA}, which allows to infer the latent codes with a small number of gradient updated ($K=3$ in all the experiments), fits particularly well the surrogate modeling task. In Figure \ref{fig:shape_reconstruction} the original shapes are reconstructed from the inferred latent codes, obtained by fitting the geometry encoder (with frozen parameters) on the SDF fields (see Fig. \ref{fig:inference}). The marching cube algorithm \cite{lorensen1998marching} is applied on the decoded SDF fields to obtain the reconstructed surface meshes. Despite some noisy regions towards the wing trailing edge and wingtip regions, it can be observed that the main geometrical features, determining the dataset shape variations, are well represented.

\section{Conclusion}
In this work, we introduced a robust methodology to build surrogate models for the prediction of aerodynamic fields in 2D and 3D on unstructured meshes that can generalize to unseen non-parametric geometric configurations.
The methodology is based on Implicit Neural Representations, enabling a continuous approximation of the target function and approximate discretization invariance. The proposed MultiScale backbone architecture allows optimal signal reconstruction, on complex fluid dynamics dataset, without extensive tuning of frequency hyperparameters.

Our experimental results on the Transonic Airfoil and XRF1 Wing datasets demonstrate the advantages of the INR-based models over state-of-the-art Graph Neural Networks (GNNs). The INR models achieve consistently the lowest mean squared error (MSE) across various aerodynamic conditions, showcasing their ability to accurately capture complex flow dynamics, including shock regions and other nonlinear phenomena.

The approximate discretization invariance of the INR models allows for training on lower resolution meshes, while maintaining high accuracy at full resolution. This leads to substantial reductions in computational cost and memory footprint.

Compared to high-fidelity CFD simulations, the surrogate model can achieve a 5 orders of magnitude speedup, allowing for real-time aerodynamic simulations. This paves the way for building surrogate models of complex 3D industrial simulations on large meshes, without substantial changes to the presented method.

\paragraph{Limitations and future work:} Despite discussing and experimentally showing the scalability of the method, the experimental section does not report results on large mesh experiments. Due to the scarce availability of open-source datasets of aircraft aerodynamics CFD simulations involving shape variations, this investigation is saved for further work. Additionally, this work only deals with the analysis of single component aerodynamic configurations (i.e. wings and airfoils), while most aircraft aerodynamic applications involve the interaction of multiple parts, such as fuselage, propellers and high-lift devices. Extending the methodology to these cases is not trivial, as it requires learning geometric representations of components with different length scales, as well as their relative position.
\begin{comment}
\section{Supplementary Material}
\begin{itemize}
   \item \textbf{Span (S)}: 90\%, 100\%, 110\%, 120\% of the base value.
    \item \textbf{Thickness (T)}: 90\%, 100\%, 110\%.
    
    \item \textbf{Dihedral Sweep (DS)}: -5 $\degree$, -2 $\degree$, 0 $\degree$, +2 $\degree$, +5 $\degree$.
\end{itemize}

\begin{itemize}
    \item \textbf{Mach Number (M)}: 0.50, 0.70, 0.74, 0.78, 0.80, 0.82, 0.84, 0.86.
    \item \textbf{Angle of Attack (α)}: -4\degree, 0\degree, 2\degree, 3\degree, 4\degree, 5\degree, 6\degree, 7\degree, 8\degree.
    \item \textbf{Daileron Angle (D)}: 0\degree, -15\degree.
\end{itemize}
\end{comment}

\section*{Acknowledgments}
This was was supported in part by the French Government represented by Agence Nationale de la Recherche (ANRT), through  CIFRE PhD Fellowship sponsored by Airbus Operations SAS and ISAE-Supaero. 

%Bibliography
\bibliographystyle{unsrt}  
\bibliography{references}  

\appendix
\section{Experimental Details}
\paragraph{Transonic Airfoil Dataset}
For each baseline model used in the experiments with the Transonic Airfoil dataset, the hyperparameters are detailed in the Table \ref{tab:end-to-end_inr},\ref{tab:vanilla_mlp},\ref{tab:gunet}. The number of epochs is chosen such that training and validation losses achieve convergence. The dataset is randomly split into the same train,validation, test (70,10,20) for all methods. The input data and output data are featurewise normalized to a standard normal distribution before being fed to the networks. The input data has four dimensions: ($x,y,M,\alpha$) while the output data has one dimension ($p$). For the end-to-end INR model, the parameters are fed through the hypernetwork while the coordinates through the main network.

\begin{table}[h]
\centering
\caption{Hyperparameters for end-to-end INR model}
\label{tab:end-to-end_inr}
\begin{tabular}{@{}lc@{}}
\toprule
\textbf{Hyperparameter}       & \textbf{Value} \\ \midrule
Layers Dimensions                                     &        [$2^{(in)}$,128,128,128,128,$1^{(out)}$]         \\
Latent Dimension                                 &       128        \\
Activation Function                          &       ReLu          \\
Encoding Type                           &       Gaussian FF     \\
N. Sampled Frequencies (per scale)                         &       64         \\
Sampling std (scales)                        &       [1,5]         \\
Hyper-network Layers Dimensions                         &       [2,128,128,128]          \\
Learning Rate                               &       2e-5         \\
Batch Size                                    &        16         \\
Epochs                                        &         5000        \\ \bottomrule
\end{tabular}
\end{table}

\begin{table}[h]
\centering
\caption{Hyperparameters for Vanilla MLP}
\label{tab:vanilla_mlp}
\begin{tabular}{@{}lc@{}}
\toprule
\textbf{Hyperparameter}       & \textbf{Value} \\ \midrule
Layers Dimensions                                    &        [$4^{(in)}$,128,128,128,128,$1^{(out)}$]        \\
Layers Type                                   &  Residual       \\
Input Encoding                            &       None     \\
Activation Function                           &       ReLu          \\
Learning Rate                                 &       2e-5         \\
Batch Size                                    &        16         \\
Epochs                                        &         5000        \\ \bottomrule
\end{tabular}
\end{table}

\begin{table}[h]
\centering
\caption{Hyperparameters for GUNET model}
\label{tab:gunet}
\begin{tabular}{@{}llc@{}}
\toprule
\textbf{Component} & \textbf{Hyperparameter}        & \textbf{Value} \\ \midrule
\multirow{2}{*}{Encoder} & Layers Dimensions                      & [$4^{(in)}$, 64, 64, 32] \\
                
                         & Activation Function              & Relu         \\ \midrule
\multirow{2}{*}{Decoder} & Layers Dimensions                      & [32, 64, 64, $1^{(out)}$]\\
                          & Activation Function              & Relu                \\
                         \midrule
\multirow{6}{*}{Processor} & Neighborhood Radius (List)  & [0.5, 1]       \\
                           & Number of levels & 2            \\
                           & Pooling Ratio       & 0.5             \\
                           & Pooling Type       & TopKPooling            \\
                           & Convolution Type    & SAGEConv         \\
                           & Max Neighbors                 & 32      \\ \midrule
\multirow{3}{*}{Shared} & Batch Size                   & 1              \\
                        & Number of Epochs             & 600            \\
                        & Learning Rate                & 0.001          \\ \bottomrule
\end{tabular}
\end{table}

\paragraph{XRF1 Wing Dataset}
For each baseline model used in the experiments with the XRF1 Wing dataset, the hyperparameters are detailed in Tables \ref{tab:encoding_inr},\ref{tab:processor_inr},\ref{tab:mgn}. For all models, featurewise normalization to the standard normal distribution, for both inputs and outputs is performed. The inputs of the INR input encoder are the 3D spatial coordinates $x,y,z$ while scalar the outputs are the SDF values. The inputs of the INR output encoder are the the 3D spatial coordinates $x$ (on the surface only) and the surface normals $n_x,n_y,n_z$, while the output are the scalar pressure values $p$.
The processor network, takes as input the input latent code $z_{in}$, concatenated with the flight parameters $M,\alpha,\delta_{ail},Re$ and produces the output latent code $z_{out}$.
For MeshGraphNet, the surface normals and the flight parameters are specified as node inputs, while the 3D relative node distances are specified on the edges.

\begin{table}[h]
\centering
\caption{Hyperparameters for INR encoder (input and output)}
\label{tab:encoding_inr}
\begin{tabular}{@{}llc@{}}
\toprule
\textbf{Component} & \textbf{Hyperparameter}        & \textbf{Value} \\ \midrule
\multirow{5}{*}{Input INR} & Layers Dimensions                   & [$3^{(in)}$, 128, 128, 128, 128,128, $1^{(out)}$] \\
                           & Latent Dimension                    & 64                       \\
                           & Sampling Std (scales)               & 1                     \\ 
                           & Hyper-network Layers Dimensions     & [128]        \\
                           & Epochs                              & 7500                     \\ \midrule
\multirow{5}{*}{Output INR} & Layers Dimensions                   & [$6^{(in)}$, 256, 256, 256, 256, 256, $1^{(out)}$] \\
                           & Latent Dimension                    & 128                       \\
                           & Sampling Std (scales)               & [1,5]                     \\ 
                           & Hyper-network Layers Dimensions     & [128]        \\
                           & Epochs                              & 1000                    \\ \midrule
\multirow{4}{*}{Shared}
                        & Learning Rate (Outer Loop)                      & 3e-5                      \\
                        & Learning Rate (Inner Loop)                      & 0.01                    \\
                        & Inner Step                                      & 3                        \\
                        & Batch Size                          & 32                       \\ \bottomrule
\end{tabular}
\end{table}

\begin{table}[h]
\centering
\caption{Hyperparameters for INR Processor}
\label{tab:processor_inr}
\begin{tabular}{@{}lc@{}}
\toprule
\textbf{Hyperparameter}       & \textbf{Value} \\ \midrule
Layers Dimensions                                    &        [$68^{(in)}$,128,128,128,$128^{(out)}$]        \\
Layers Type                                   &  Residual       \\
Activation Function                           &       Silu          \\
Learning Rate                                 &       5e-6         \\
Batch Size                                    &        128         \\
Epochs                                        &         1000        \\ \bottomrule
\end{tabular}
\end{table}

\hfill
\begin{table}[h]
\centering
\caption{Hyperparameters for MeshGraphNet model}
\label{tab:mgn}
\begin{tabular}{@{}llc@{}}
\toprule
\textbf{Component} & \textbf{Hyperparameter}        & \textbf{Value} \\ \midrule
\multirow{2}{*}{Node Encoder} & Layers Dimensions                      & [$7^{(in)}$, 64, 64, 64] \\
                
                         & Activation Function              & Relu         \\ \midrule
\multirow{2}{*}{Edge Encoder} & Layers Dimensions                      & [3, 64, 64, 64] \\
                
                         & Activation Function              & Relu         \\ \midrule
\multirow{2}{*}{Decoder} & Layers Dimensions                      & [64, 64, 64, $1^{(out)}$]\\
                          & Activation Function              & Relu                \\
                         \midrule
\multirow{1}{*}{Processor}
                           & Number of blocks & 12            \\ \midrule
\multirow{3}{*}{Shared} & Batch Size                   & 4              \\
                        & Number of Epochs             & 150            \\
                        & Learning Rate                & 0.001          \\ \bottomrule
\end{tabular}
\end{table}

\section{Additional Results}

\subsection{Transonic Airfoil Dataset}

\begin{figure}[h!]
    \centering
    \begin{subfigure}[b]{0.5\textwidth}
        \centering
        \adjustbox{valign=c}{\includegraphics[width=\linewidth]{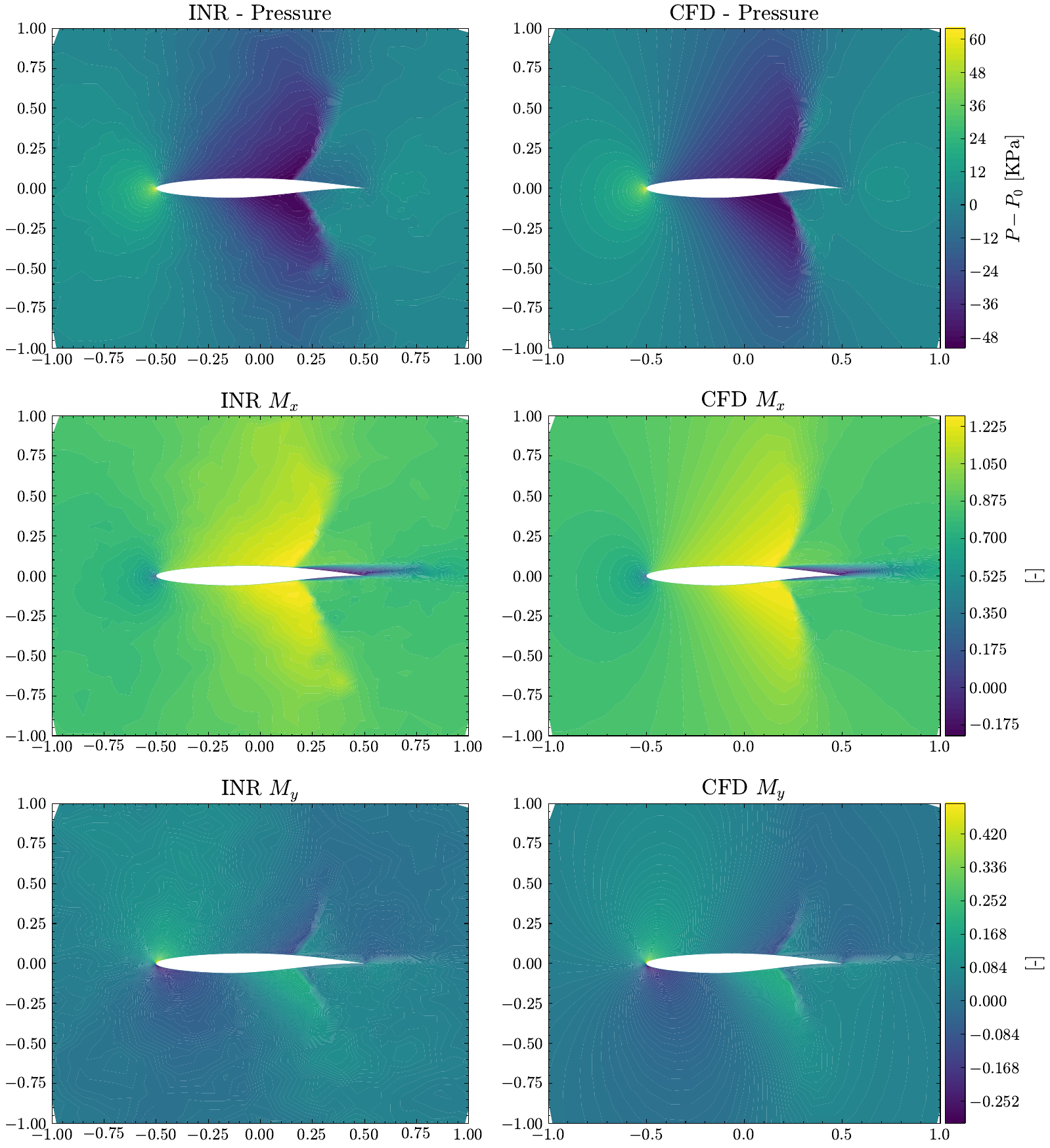}}
        \caption{$\alpha=2.04$,$M=0.86$}
        \label{fig:first_image}
    \end{subfigure}%
    \hfill
    \begin{subfigure}[b]{0.5\textwidth}
        \centering
        \adjustbox{valign=c}{\includegraphics[width=\linewidth]{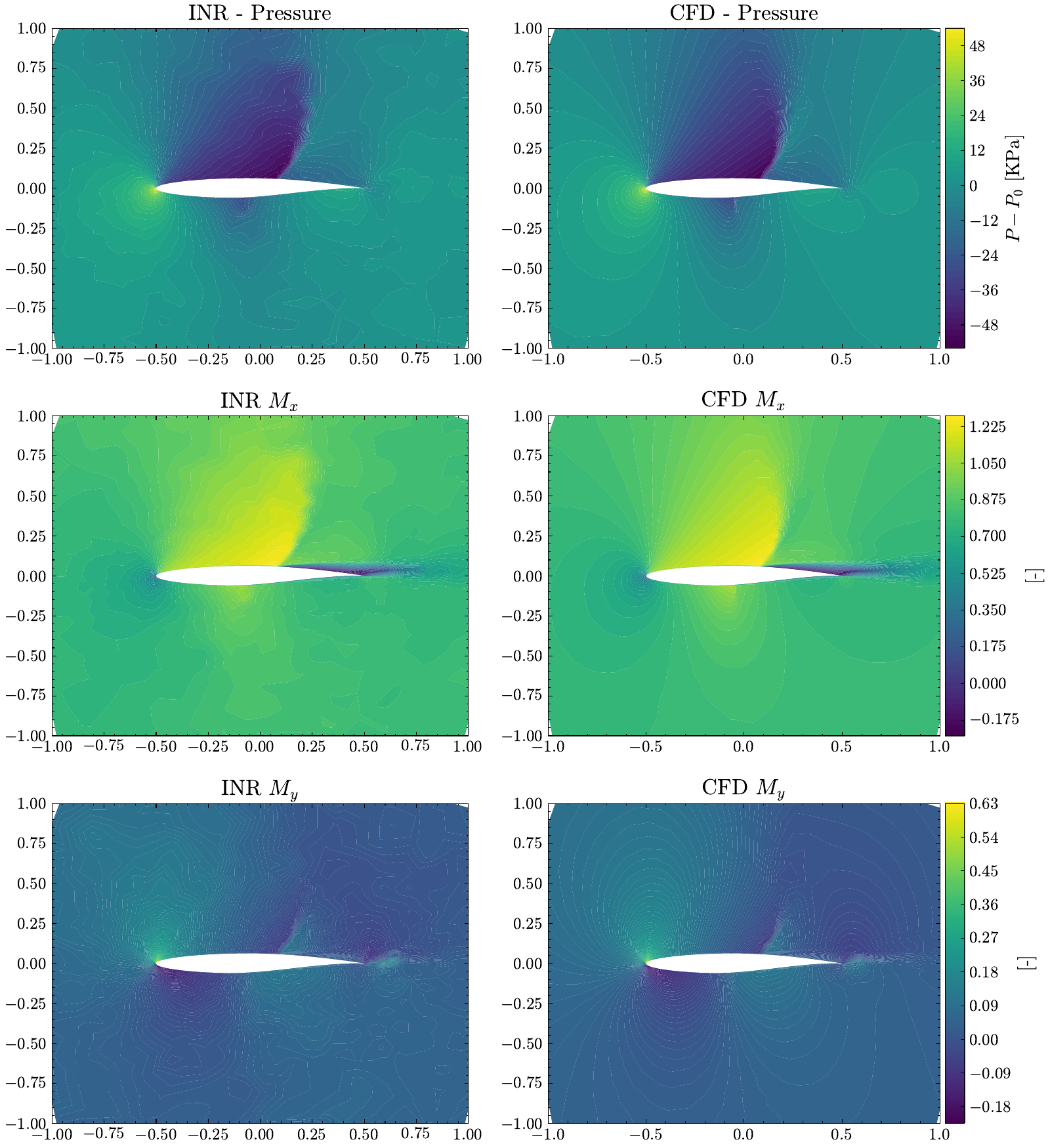}}
        \caption{$\alpha=3.45$,$M=0.80$}
        \label{fig:second_image}
    \end{subfigure}%
    \caption{\textbf{Top:} Pressure Distribution. \textbf{Center:} Horizontal Local Mach number $M_x$ (Horizontaly velocity normalized by the speed of sound). \textbf{Bottom:} Vertical Local Mach Number $M_y$.}
    \label{fig:mutli_output}
\end{figure}

\paragraph{Multi-Output INR} 
The proposed INR model can be used to perform multi-output prediction of physical quantities. This is particularly relevant for the definition of physics based losses, as governing laws typically relate multiple physical quantities. In \ref{fig:mutli_output} an end-to-end INR surrogate model is trained to output the pressure field and the two-dimensional velocity vector field around the RAE2822 airfoil. The predictions are in good accordance with the reference CFD targets. The choice of the frequency embedding can be more challenging in this case, as different physical variables can have different frequency characteristics: pressure fields are smoother and do not present sharp gradients across the boundary layer and in the wake flow, compared to velocity fields. This is an additional justification of the flexibility of the proposed MultiScale INR architecture, as multiple frequency encoding can balance the reconstruction accuracy of signals with different bandwidths. 

\begin{figure}
  \centering
  \includegraphics[width=1\linewidth]{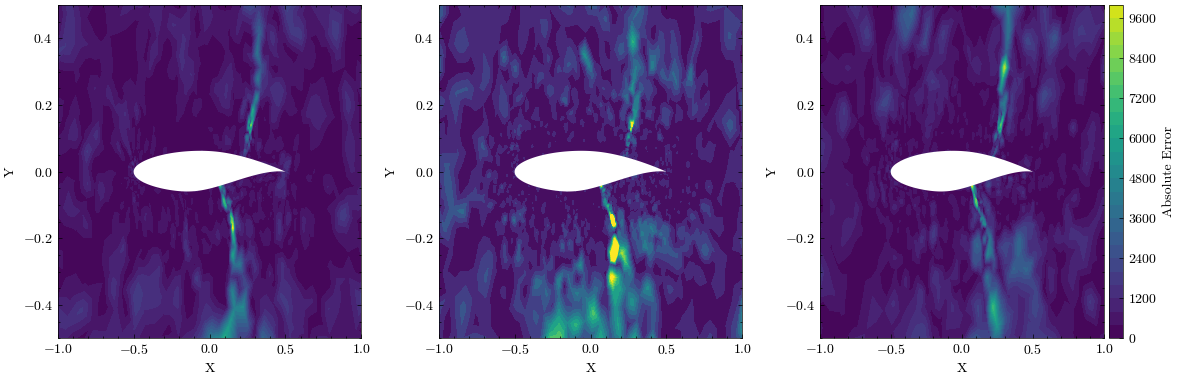}
  \caption{Contour plot of the magnitude of the prediction error using different values of $\sigma$. \textbf{Left:} $\sigma=1$ at $\alpha=0 [deg]$,$M=0.84$. \textbf{Center:} $\sigma=5$. \textbf{Right:} $\sigma=[1,5]$.}
  \label{fig:sigma_comparison}
\end{figure}

\paragraph{Effect of encoding scale}
The parameter $\sigma$ controlling the range of sampled embedding frequencies can have an important influence on the network reconstructions, namely in terms of underfitting and overfitting behaviour of the model. This was more in detail discussed in Section \ref{sec:methodology} and briefly shown in the 1D regression example of Fig. \ref{fig:both_subfigures}. Here, we complement the discussion by presenting a typical error distribution plot (Fig. \ref{fig:sigma_comparison}) for the model predictions performed with different choices of the $\sigma$ parameter: $\sigma=1$, $\sigma=5$ and a Multiscale Architecture with $\sigma=[1,5]$. Is is clear that the reconstructions for higher values of $\sigma$ become noisy, as the kernel is not filtering out the spurious high-frequency components present in the training dataset.
The result with $\sigma=1$ presents more localized error peaks in proximity of the shock region, where sharp variation of the pressure fields are present. The multiscale architecture appears to mediate the effect of the two different scale embeddings, by producing slightly lower error peaks in the strongly non-linear regions and removing the noisy patterns introduced by the larger kernel bandwidth.

\paragraph{POD and GPR} Additional baseline comparison with a POD+GPR methodology are shown. This surrogate combines Proper Orthogonal Decomposition (POD) with Gaussian Process Regression (GPR) to perform dimensionality reduction via linear mode decomposition and regression in the low-dimensional space of POD coefficients. This is a traditional surrogate model, widely used for fluid dynamics application, thanks to the capability of preserving the more salient features in the training dataset, and to respect physical constraints such as Dirichlet boundary conditions and conservation of mass \cite{berkooz1993proper}. In this study, the truncated Singular Value Deomposition of the snapshot matrix containing the training samples is performed, keeping the first 50 modes. A Gaussian Process Regressor, mapping the input parameters to the reduced coefficients, is implemented in the SMT  \cite{saves2024smt} with a Matérn kernel. Due to the fixed geometry and resolution constraints of the POD formulation, this model offline and online stage are carried out on the same full mesh.
In Table \ref{tab:comparison_rae_pod} a summary of the surrogate model metrics on the RAE2822 dataset is provided. Overall, the method achieves error metrics which are competitive with state of the art model (as GNNs), on problems with fixed grid settings.

\begin{figure}
  \centering
  \includegraphics[width=1\linewidth]{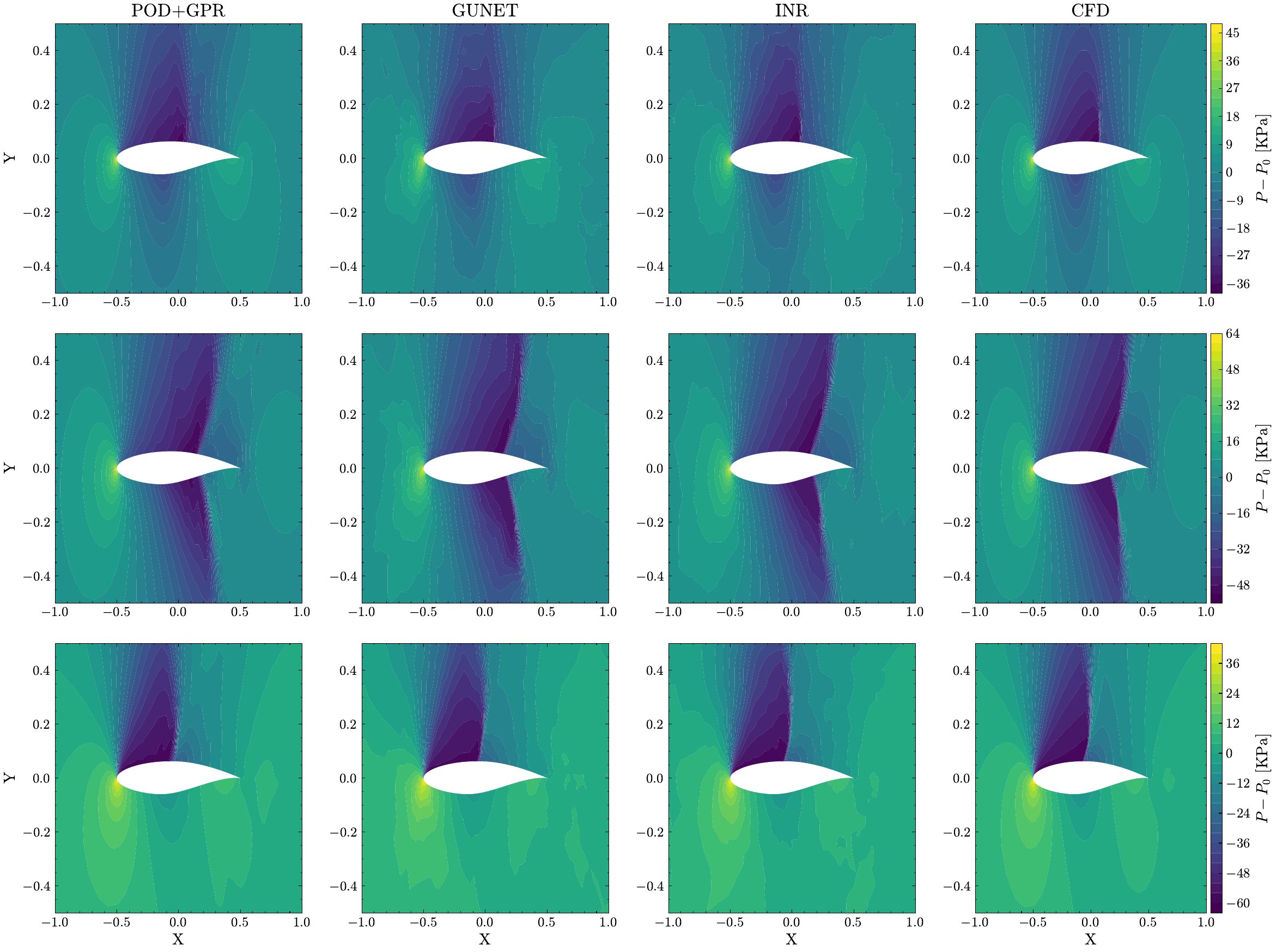}
  \caption{Pressure field distribution prediction with different surrogate models and comparison with CFD. \textbf{Top:} Angle of Attack [0.75 deg], Mach Number [0.74]. \textbf{Center:} Angle of Attack [2.78 deg], Mach Number [0.85]. \textbf{Bottom:} Angle of Attack [5.58 deg], Mach Number [0.72].}
  \label{fig:contour_pod}
\end{figure}

\begin{figure}
  \centering
  \includegraphics[width=1\linewidth]{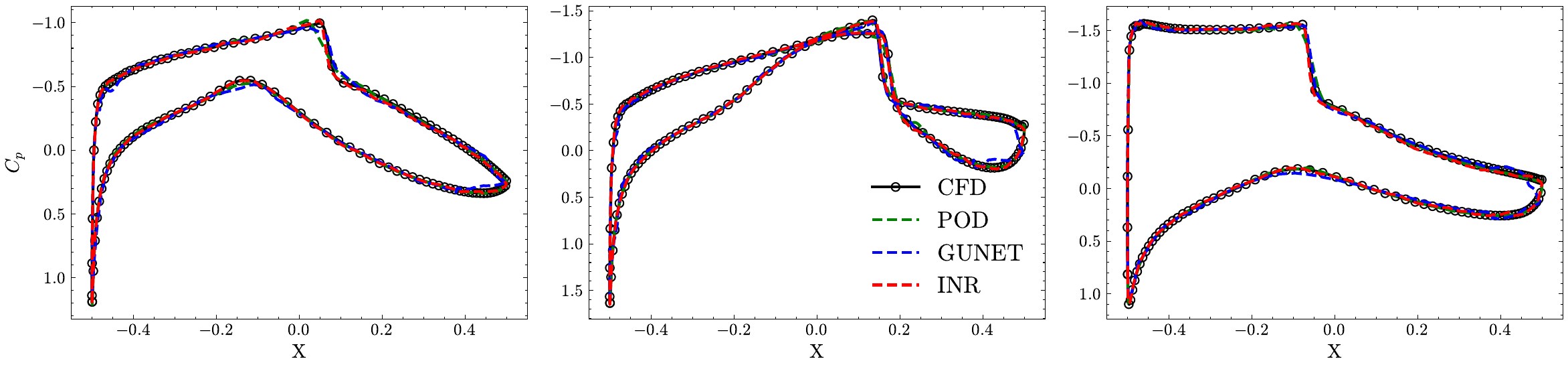}
  \caption{Surface Pressure Coefficient distribution prediction with different surrogate models and comparison with CFD. \textbf{Left:} Angle of Attack [0.75 deg], Mach Number [0.74]. \textbf{Center:} Angle of Attack [2.78 deg], Mach Number [0.85]. \textbf{Right:} Angle of Attack [5.58 deg], Mach Number [0.72].}
  \label{fig:cp_pod}
\end{figure}

While providing physically plausible results away from the shock region, most of the model error are located in proximity of the pressure jump, mainly due to the linear formulation of the method, as visible in Figure \ref{fig:contour_pod} and Figure \ref{fig:cp_pod}, where the POD-GPR sligthly oversmooth the shock discontinuity. Not surprisingly, this method can be efficiently trained in around few minutes: this is a big advantage that makes it particularly attractive for rapid surrogate modeling development. 
We note here, that we did not try to experiment with more advance POD and GPR formulations, namely clustered POD \cite{dupuis2018aerodynamic,catalani2022machine}.It is reasonable to assume that ad-hoc modification of the linear decomposition formulation can improve the method performance. However, this does not hold much relevance for the scope of the current study.

\begin{table}[h]
\centering
\caption{Performance comparison of different models in terms of error (MSE), training time (s), and inference time (ms).}
\label{tab:comparison_rae_pod}
\begin{tabular}{@{}lccc@{}}
\toprule
\textbf{Model}      & \textbf{MSE} & \textbf{Training Time (s)} & \textbf{Inference Time (ms)} \\ \midrule
INR          &   0.0020    &     9497   &       4              \\
POD+GPR       &   0.0040    &     260  &       10                 \\ \bottomrule
\end{tabular}
\end{table}

\end{document}